\documentstyle[12pt]{article}

\begin{document}
\parskip 1ex
\setcounter{page}{1}
\oddsidemargin 0pt
\evensidemargin 0pt
\topmargin -40pt
%
\newcommand{\be}{\begin{equation}}
\newcommand{\ee}{\end{equation}}
\newcommand{\beq}{\begin{eqnarray}}
\newcommand{\eeq}{\end{eqnarray}}
\def\a{\alpha}
\def\b{\beta}
\def\g{\gamma}
\def\G{\Gamma}
\def\d{\delta}
\def\e{\epsilon}
\def\z{\zeta}
\def\h{\eta}
\def\th{\theta}
\def\k{\kappa}
\def\l{\lambda}
\def\L{\Lambda}
\def\m{\mu}
\def\n{\nu}
\def\x{\xi}
\def\X{\Xi}
\def\p{\pi}
\def\P{\Pi}
\def\r{\rho}
\def\s{\sigma}
\def\S{\Sigma}
\def\t{\tau}
\def\f{\phi}
\def\F{\Phi}
\def\c{\chi}
\def\w{\omega}
\def\W{\Omega}
\def\de{\partial}

\def\pct#1{(see Fig. #1.)}

\begin{titlepage}
\hbox{\hskip 12cm CERN-TH/97-306  \hfil}
\hbox{\hskip 12cm UCLA/97/TEP/24 \hfil}
\hbox{\hskip 12cm ROM2F-97/37  \hfil}
\hbox{\hskip 12cm hep-th/9711059 \hfil}
\vskip 0.8cm
\begin{center}  {\Large  \bf  Tensor \ and \ Vector \ Multiplets \vskip .6cm in \
Six-Dimensional
\ Supergravity}

\vspace{0.8cm}
 
{\large \large Sergio Ferrara$^{\dagger, \ddagger}$, \ \ Fabio 
Riccioni$^{*,}\footnote{I.N.F.N. Fellow}$ 
\ \ and \ \ Augusto Sagnotti$^{*}$}
\vspace{0.6cm}

{$^{\dagger}$ \sl Theory \ Division, \ \ CERN  \\ CH-1211  Geneva 23, \ \  SWITZERLAND}

\vspace{0.3cm}

{$^{\ddagger}$  Department of Physics and Astronomy,\\ University of California, Los
Angeles, CA 90024 \ U.S.A.}

\vspace{0.3cm}

{$^{*}$ \sl Dipartimento di Fisica, \ \ Universit{\`a} di Roma \ ``Tor Vergata'' \\
I.N.F.N.\ - \ Sezione di Roma \ ``Tor Vergata'', \\ Via della Ricerca Scientifica , 1 \
\ \ 00133 \ Roma \ \ ITALY}
\end{center}
\vskip .8cm

\abstract{We construct the complete coupling of $(1,0)$ supergravity in six dimensions
to $n$ tensor multiplets, extending previous results to all orders in the fermi fields. 
We then add couplings to vector multiplets, as dictated by the generalized Green-Schwarz
mechanism.  The resulting theory embodies factorized gauge and supersymmetry anomalies,
to be disposed of by fermion loops, and is determined by corresponding Wess-Zumino
consistency conditions, aside from a quartic coupling for the gaugini.  The
supersymmetry algebra contains a corresponding extension that plays a crucial role for
the consistency of the construction. We leave aside gravitational and mixed anomalies,
that would only contribute to higher-derivative couplings.}
\vfill
\hbox{\hskip 1.2cm CERN-TH/97-306  \hfil}
\hbox{\hskip 1.2cm November 1997 \hfil}
\end{titlepage}
\makeatletter
\@addtoreset{equation}{section}
\makeatother
\renewcommand{\theequation}{\thesection.\arabic{equation}}
\addtolength{\baselineskip}{0.3\baselineskip} 

\section{Introduction}

One of the most striking features of perturbative superstring theory in ten  dimensions
is the absence of anomalies. In the type-IIB theory this is  realized by miraculous
cancellations between various contributions \cite{agw}, while in the type-I and
heterotic theories the  Green-Schwarz mechanism \cite{gs} generates anomalous couplings
that exactly cancel the contributions of fermion loops, once one restricts the gauge
group to be $SO(32)$ for the type I theory  and $SO(32)$ or
$E_8 \times E_8$ for the heterotic theory. All these $N=1$ theories  are very
interesting, since they can be naturally compactified to  rich spectra of $N=1$ theories
in four dimensions. In this context, an interesting intermediate step is the study of
$(1,0)$  vacua in six dimensions, since in these compactifications the absence of 
anomalies is a strong restriction on the low-energy physics.

The massless representations of $(1,0)$ supersymmetry in six dimensions,  labeled by
their
$SU(2)\times SU(2)$ representations, are the gravity multiplet $\bigl(
(1,1)+2(1,\frac{1}{2})+(1,0) \bigr)$, the tensor multiplet $\bigl(
(0,1)+2(0,\frac{1}{2})+(0,0)
\bigr)$, the vector multiplet $\bigl( (\frac{1}{2},\frac{1}{2})+2(\frac{1}{2},0)
\bigr)$, and the hypermultiplet $\bigl( 2(0,\frac{1}{2})+4(0,0) \bigr)$.
\cite{dfr} considered pure $(1,0)$ supergravity, and in
\cite{romans} $(1,0)$ supergravity was coupled to an arbitrary number of tensor 
multiplets to lowest order in the fermi fields, while \cite{ns} considered the case of a
single tensor multiplet and an arbitrary number of  hypermultiplets. It was then found
\cite{as} that the model  in
\cite{romans} can be coupled to Yang-Mills multiplets in a way determined by the
residual gauge and gravitational anomalies. The relation to the supersymmetry anomaly
was elucidated in
\cite{fms}, to lowest order in the fermi fields, and the resulting coupling to 
hypermultiplets was then partly constructed in \cite{ns2}.

Letting $n_T$, $n_V$ and $n_H$ denote the numbers of tensor, vector and hypermultiplets,
the condition that the term $tr R^4$ be absent  in the anomaly polynomial \cite{r-d},
\be n_H - n_V +29 n_T = 273 \quad ,
\ee allows a large number of possible vacua. Perturbative heterotic vacua in six
dimensions can be obtained by orbifold compactifications or by compactifications on
smooth $K3$ manifolds with instanton backgrounds. Anomaly cancellation requires that the
total  instanton number be 24, and these vacua include a single tensor multiplet, as one
can easily see reducing the ten dimensional  low-energy theory.  The situation is quite
different in perturbative six-dimensional  type I vacua since, as suggested in
\cite{cargese}, these models are  determined  by a parameter space orbifold
(orientifold) construction, and this naturally allows several tensor multiplets \cite{bs}.
The residual anomaly  polynomial in general does not factorize, and several
antisymmetric tensors contribute  to the cancellation in a generalized Green-Schwarz 
mechanism \cite{as}. Moreover, the low-energy supergravity exhibits singularities  in
the moduli space of tensor multiplets, corresponding to infinite  gauge coupling
constants \cite{as}. These singularities have attracted some interest, since they
reflect the  presence in the vacuum of string excitations with vanishing tension
\cite{sw}, and signal a new kind of phase transition \cite{dmw}. The conjectured type I
- heterotic duality \cite{wit} relates  these peculiar perturbative type-I vacua to
corresponding non-perturbative heterotic vacua. 

In this paper we construct the complete $(1,0)$ supergravity coupled to tensor and vector
multiplets.  This theory contains (reducible) gauge and supersymmetry anomalies induced
by tensor couplings, that here are  completely determined solving Wess-Zumino
consistency conditions.  In Section 2  we construct minimal supergravity coupled to $n$
tensor multiplets, thus completing
\cite{dfr,romans} to all orders in the fermi fields. In Section 3 we include all
additional couplings to vector multiplets, thus  completing
\cite{as,fms}.  Some of the higher-order fermion couplings were previously introduced in
\cite{ns2}, where couplings to hypermatter were also considered.  Section 4 is devoted
to a discussion of our results, while the Appendix  collects some details on our
notation and a number of identities  used in our derivations.  While this work was being
typed, a lagrangian superspace formulation of the  theory with tensor multiplets
only was presented in \cite{padova}.

\section{Minimal Supergravity in Six Dimensions Coupled to $n$ Tensor  Multiplets}

In this Section we describe minimal $(1,0)$  six-dimensional supergravity coupled to $n$
tensor multiplets.   Simple supersymmetry in six dimensions is generated by an $Sp(2)$
doublet of chiral spinorial charges $Q^a$ $(a=1,2)$, obeying the symplectic Majorana
condition
\be Q^a = \e^{ab} C \bar{Q}^T_b \quad ,
\ee where $\e^{ab}$ is the $Sp(2)$  antisymmetric invariant tensor. Since all  fermi
fields appear as $Sp(2)$ doublets, from now on we will mostly use $\Psi$ to denote a
doublet 
$\Psi^a$.  Further details on this notation   may be found in the Appendix.

Let us begin by reviewing the work of Romans \cite{romans}. The  theory includes the
vielbein
$e_\m{}^a$, a left-handed gravitino $\Psi_\m$, $(n+1)$  antisymmetric tensors
$B^r_{\m\n}$ $(r=0,...,n)$  obeying (anti)self-duality conditions, $n$ right-handed ``tensorini''
$\chi^m$  $(m=1,...,n)$, and $n$ scalars. The scalars parameterize the coset space
$SO(1,n)/SO(n)$, and are  thus associated to the $SO(1,n)$ matrix $(r=0,...n)$
\be V =\pmatrix{v_r \cr x^m_r}\quad ,
\ee whose matrix elements satisfy the constraints
\beq & & v^r v_r =1 \quad , \nonumber\\ & & v_r v_s - x^m_r x^m_s = \eta_{rs}\quad ,
\nonumber \\ & & v^r x^m_r =0 \quad . \label{scalars}
\eeq Defining
\be G_{rs} = v_r v_s + x^m_r x^m_s \quad ,
\ee the tensor (anti)self-duality conditions can be succinctly written
\be G_{rs} H^{s \m\n\r} =\frac{1}{6e} \e^{\m\n\r\a\b\g} H_{r \a\b\g}\quad ,
\label{selfdual}
\ee where $H^r_{\m\n\r}= 3 \de_{[\m} B^r_{\n\r ]}$. These relations only hold to lowest
order in the fermi fields, and imply that  $v_r H^r_{\m\n\r}$ is self dual, while the
$n$ tensors $x^m_r H^r_{\m\n\r}$ are antiself dual, as one can see using eqs.
(\ref{scalars}). The divergence of eq. (\ref{selfdual})  yields the second-order tensor
equation
\be D_\m (G_{rs} H^{s\m\n\r} )=0  
\label{tensoreq}
\ee while, to lowest order, the fermionic equations are
\be
\g^{\m\n\r} D_\n \Psi_\r + v_r H^{r \m\n\r} \g_\n \Psi_\r-
\frac{i}{2} x^m_r H^{ r \m\n\r} \g_{\n\r} \chi^m + 
\frac{i}{2} x^m_r \de_\n v^r \g^\n \g^\m \chi^m =0 
\ee and
\be
\g^\m D_\m \chi^m -\frac{1}{12} v_r H^{r \m\n\r} \g_{\m\n\r} \chi^m -\frac{i}{2} x^m_r
H^{r
\m\n\r} \g_{\m\n} \Psi_\r-
\frac{i}{2} x^m_r \de_\n v^r \g^\m \g^\n \Psi_\m =0\quad. 
\ee Varying the fermi fields in them with the supersymmetry transformations 
\beq 
& & \delta e_\m{}^a = -i ( \bar{\e} \g^a \Psi_\m ) \quad , \nonumber\\ 
& & \delta B^r_{\m\n} =i v^r ( \bar{\Psi}_{[\m} \g_{\n]} \e )
+\frac{1}{2} x^{mr} ( \bar{\chi}^m \g_{\m\n} \e ) \quad , \nonumber\\ 
& & \delta v_r = x^m_r ( \bar{\e} \chi^m )\nonumber\\
& & \delta \Psi_\m = D_\m \e +\frac{1}{4} v_r H^r_{\m\n\r} \g^{\n\r} \e \quad , 
\nonumber\\ & & \delta \chi^m =\frac{i}{2} x^m_r
\de_\m v^r \g^\m \e +\frac{i}{12} x^m_r H^r_{\m\n\r} \g^{\m\n\r} \e \quad ,  \label{susy}
\eeq 
generates the bosonic equations, using also eqs. (\ref{selfdual}) and
(\ref{tensoreq}). Thus, the scalar field equation is
\be x^m_r D_\m (\de^\m v^r ) +\frac{2}{3} x^m_r v_s H^r_{\a\b\g}  H^{s \a\b\g} =0 \quad ,
\ee while the Einstein equation is
\be R_{\m\n} -\frac{1}{2} g_{\m\n} R + \de_\m v^r \de_\n v_r -
\frac{1}{2} g_{\m\n} \de_\a v^r \de^\a v_r - G_{rs} H^r_{\m\a\b} H^s{}_\n{}^{\a\b}
=0\quad.
\ee 
To this order, this amounts to a proof of  supersymmetry,  and it is also possible
to show that the  commutator of two supersymmetry transformations on the bosonic fields
closes on the local symmetries:
\beq 
& & [ \delta_1 , \delta_2 ] = {\delta}_{gct}( \xi^\m = -i ({\bar{\e}}_1 
\g^\m \e_2 )) +
\delta_{tens} (\L^r_\m = -\frac{1}{2} v^r \xi_\m -\xi^\n B^r_{\m\n}) 
\nonumber \\ & & +\delta_{SO(n)}(A^{mn} = \xi^\m x^{mr}(\de_\m x^n_r ) ) 
+\delta_{Lorentz} (\W^{ab} =-\xi_\m (\w^{\m a b} - v_r H^{r \m a b})) \quad
.\label{susyalg}
\eeq 
To this order, one can not  see the local supersymmetry transformation in the gauge
algebra, since the expected parameter,  $\xi^\m \Psi_\m$, is generated  by bosonic
variations. As usual, the spin connection satisfies its equation of motion, that to
lowest order in the fermi  fields is 
\be D_\m e_\n{}^a - D_\n e_\m{}^a =0\quad ,
\ee and implies the absence of torsion.

Completing these equations will require terms cubic in the fermi fields in the fermionic
equations, and  terms quadratic in the fermi fields in their supersymmetry
transformations.  Supersymmetry will then determine corresponding modifications of the 
bosonic equations, and the (anti)self-duality conditions 
(\ref{selfdual}) will also be
modified by terms quadratic in the fermi fields. Supercovariance actually   fixes  all
terms containing the gravitino in the first-order equations and in the supersymmetry
variations of fermi fields.

The supercovariant forms
\be
\hat{\w}_{\m\n\r} = \w^0_{\m\n\r}  -\frac{i}{2} (\bar{\Psi}_\m \g_\n \Psi_\r 
+\bar{\Psi}_\n \g_\r
\Psi_\m +\bar{\Psi}_\n \g_\m \Psi_\r )\quad ,
\ee
\beq & & \hat{H}^r_{\m\n\r} = H^r_{\m\n\r}  -\frac{1}{2} x^{mr} ( \bar{\chi}^m \g_{\m\n}
\Psi_\r +
\bar{\chi}^m 
\g_{\n\r}\Psi_\m + \bar{\chi}^m \g_{\r\m} \Psi_\n )  \nonumber\\ & & - \frac{i}{2} v^r
(\bar{\Psi}_\m \g_\n \Psi_\r +\bar{\Psi}_\n \g_\r
\Psi_\m +\bar{\Psi}_\r \g_\m \Psi_\n ) \quad ,
\eeq
\be
\hat{\de_\m v^r} = \de_\m v^r -x^{mr} (\bar{\chi}^m \Psi_\m )\quad ,
\ee where
\be
\w^0_{\m\n\r} =\frac{1}{2} e_{\r a}(\de_\m e_\n{}^a -\de_\n e_\m{}^a ) -\frac{1}{2} e_{\m
a}(\de_\n e_\r{}^a -\de_\r e_\n{}^a) +\frac{1}{2} e_{\n a} (\de_\r e_\m{}^a - \de_\m
e_\r{}^a )
\ee is the standard spin connection in the absence of  torsion, do not generate
derivatives of the  parameter under supersymmetry. In the same spirit, one can consider
the supercovariant transformations 
\beq & & \delta \Psi_\m = \hat{D}_\m \e +\frac{1}{4} v_r \hat{H}^r_{\m\n\r} 
\g^{\n\r} \e \quad , \nonumber \\ & & \delta \chi^m = \frac{i}{2} x^m_r (\hat{\de_\m
v^r} ) \g^\m
\e +\frac{i}{12} x^m_r \hat{H}^r_{\m\n\r} \g^{\m\n\r} \e \quad . \label{newsusy}
\eeq The tensorino transformation is complete, while the gravitino  transformation could
include additional terms quadratic in the tensorini. On the other hand, one does not
expect modifications of  the bosonic transformations in the complete theory.

\subsection{Complete Supersymmetry Algebra}

The algebra (\ref{susyalg}) has been obtained varying only the fermi fields in the
bosonic supersymmetry transformations. The next step is to compute  the commutator
varying the bosonic fields as well.  There is no important novelty in the complete
commutator on $v^r$ and on the vielbein $e_\m{}^a$. However, the local Lorentz parameter
is modified and takes the form
\be
\W^{ab}= -\xi^\m (\hat{\w}_\m{}^{ab} -v_r \hat{H}^r_\m{}^{ab})\quad 
\ee while, as anticipated, the supersymmetry parameter is
\be
\zeta = \xi^\m \Psi_\m \quad .
\ee These results are obtained using the torsion equation for $\hat{\w}$,
\be
\hat{D}_\m e_\n{}^a - \hat{D}_\n e_\m{}^a = 2S^a{}_{\m\n}= -i(\bar{\Psi}_\m
\g^a \Psi_\n ) \quad .
\ee

One can also compute the commutator on $x^m_r$. Eqs. (\ref{scalars}) determine its
supersymmetry variation 
\be
\delta x^m_r = v_r (\bar{\e} \chi^m ) \quad ,
\ee and the resulting commutator includes a  local $SO(n)$ transformation of parameter
\be A^{mn} =\xi^\m x^{mr} (\de_\m x^n_r ) +(\bar{\chi}^m \e_2 )(\bar{\chi}^n
\e_1 ) -(\bar{\chi}^m \e_1 )(\bar{\chi}^n \e_2 ) \quad .
\ee

New results come from the complete commutator on $B^r_{\m\n}$, where one needs to use the
(anti)self-duality conditions. Supercovariantization is at work here, since these
conditions are first-order equations, that become
\be  G_{rs} \hat{H}^s_{\m\n\r} =\frac{1}{6e} \e_{\m\n\r\a\b\g} \hat{H}_r^{\a\b\g}
\quad .\label{selfdual2}
\ee It is actually possible to alter these conditions demanding that the modified tensor
\be
\hat{\cal H}^{r}_{\m\n\r}=
\hat{H}^r_{\m\n\r} +i\a v^r (\bar{\chi}^m \g_{\m\n\r}\chi^m )\label{newtensor}
\ee satisfy (anti)self-duality conditions as in eq. (\ref{selfdual2}).  Using eqs.
(\ref{scalars}), one can see that the new $\chi^2$ terms contribute only to the
self-duality condition, while the tensors
$x^m_r \hat{H}^r_{\m\n\r}$ remain antiself dual without extra $\chi^2$  terms.
Consequently,  since the commutator on $B^r_{\m\n}$  uses only the antiself-duality
conditions, the result does not contain  terms proportional to $\a$. The commutator on
the tensor fields  generates all local symmetries in the proper form, aside from the 
extra terms
\be [\delta_1 , \delta_2 ]_{extra} B^r_{\m\n} =\frac{1}{2}v^r (\bar{\e}_1 \chi^m
)(\bar{\chi}^m
\g_{\m\n} \e_2 ) -\frac{1}{2} v^r (\bar{\e}_2 \chi^m )(\bar{\chi}^m \g_{\m\n} \e_1
)\quad ,
\ee that may be canceled adding $\chi^2$ terms to the transformation
of the gravitino. The
most general expression one can add is
\be
\delta^\prime \Psi_\m = ia \ \g_\m \chi^m (\bar{\e} \chi^m )+  i b \ \g_\n \chi^m (\bar{\e}
\g_\m{}^\n \chi^m )  + i c \ \g_{\m\n\r} \chi^m (\bar{\e} \g^{\n\r} \chi^m ) \quad ,
\ee with $a$, $b$ and $c$ real coefficients, and the total commutator on $B^r_{\m\n}$
then leads to the relations
\be a + b = -\frac{1}{2} \quad ,\qquad b + 2c = 0 \quad .
\ee The commutator on $e_\m{}^a$ now closes with a local Lorentz parameter modified by
the addition of
\be
\Delta{\W}^{ab}  =-\frac{1}{2}[(\bar{\chi}^m \e_1 )(\bar{\e}_2 \g^{ab} \chi^m)
-(\bar{\chi}^m \e_2 )(\bar{\e}_1 \g^{ab} \chi^m )] \quad , 
\ee while the commutators on the scalar fields are not modified.

One can now start to compute the commutators on fermi fields, that as usual close only
on shell. Following \cite{schwarz}, we will actually use this result to  derive the
complete fermionic equations. Let us begin with the  commutator on the tensorini, using
eq. (\ref{newsusy}). This fixes the free parameter in the gravitino variation and the
parameter $\a$ in eq. (\ref{newtensor}), so that
\be a = -\frac{3}{8} \quad , \qquad  b = -\frac{1}{8} \quad , 
\qquad c = \frac{1}{16} \quad , \qquad \a = -\frac{1}{8} \quad .
\ee Supercovariance determines the field equation of the tensorini up to a term
proportional to
$\chi^3$. Closure of  the algebra fixes this additional term, and the end result is
\beq & &\g^\m \hat{D}_\m \chi^m -\frac{1}{12}v_r \hat{H}^r_{\m\n\r} \g^{\m\n\r}
\chi^m -\frac{i}{2} x^m_r \hat{H}^{r \m\n\r} \g_{\m\n} \Psi_\r \nonumber \\ &
&-\frac{i}{2} x^m_r ( \hat{\de_\n v^r} ) \g^\m \g^\n \Psi_\m  -\frac{i}{2} \g^\a \chi^n
(\bar{\chi}^n \g_\a \chi^m ) = 0
\quad .\label{chieq}
\eeq The complete commutator of two supersymmetry transformations on the tensorini is
then
\be [\delta_1 ,\delta_2 ] \chi^m = \delta_{gct} \ \chi^m +\delta_{Lorentz} \ \chi^m
+\delta_{SO(n)} \ \chi^m +\delta_{susy} \ \chi^m
 + \frac{1}{4} \g^\a \xi_\a \ [{\rm eq.} \ \chi^m ] \quad .
\ee A similar result can be obtained for the gravitino. In this case the complete
equation,
\beq & & \g^{\m\n\r} \hat{D}_\n \Psi_\r +\frac{1}{4} v_r \hat{H}^r_{\n\a\b}
\g^{\m\n\r} \g^{\a\b} \Psi_\r -\frac{i}{2}x^m_r \hat{H}^{r \m\n\r} \g_{\n\r}
\chi^m  +\frac{i}{2} x^m_r (\hat{\de_\n v^r}) \g^\n \g^\m \chi^m \nonumber \\ & &
+\frac{3i}{2}
 \g^{\m\a} \chi^m  (\bar{\chi}^m \Psi_\a ) -\frac{i}{4} \g^{\m\a} \chi^m (\bar{\chi}^m
\g_{\a\b}
\Psi^\b ) +\frac{i}{4} \g_{\a\b} \chi^m (\bar{\chi}^m \g^{\m\a} \Psi^\b ) \nonumber\\ &
& -
\frac{i}{2} \chi^m (\bar{\chi}^m \g^{\m\a} \Psi_\a ) =0 \quad ,
\label{gravitinoeq}
\eeq is fixed by  supercovariance, and the commutator closes up to  terms proportional
to a particular combination of eq. (\ref{gravitinoeq}) and its $\g$-trace. Moreover,  a
non-trivial symplectic structure makes its first appearance in  a commutator, so that
the final result is
\beq & & [\delta_1 ,\delta_2 ] \Psi_\m^a = \delta_{gct} \Psi_\m^a
+\delta_{Lorentz}\Psi_\m^a +\delta_{susy}\Psi_\m^a \nonumber\\ & & +\frac{3}{8} \xi^\a
\g_\a ([eq.\  \Psi_\m ] -\frac{1}{4}
\g_\m [\g -trace ])^a \nonumber \\ & & +\frac{1}{96} \sigma^i_b{}^a \g^{\a\b\g}
\xi^i_{\a\b\g} ([eq.\  \Psi_\m ] -\frac{1}{4} \g_\m [\g -trace ])^b \quad ,
\label{gravitinoalg}
\eeq where
\be
\xi^i_{\a\b\g} = -i [\bar{\e}_1 \g_{\a\b\g} \e_2 ]^i \quad . \label{anomalousxi}
\ee

Summarizing, from the algebra we have obtained the complete fermionic equations of
$(1,0)$ six-dimensional supergravity coupled to $n$ tensor multiplets. In addition, the
modified 3-form
\be
\hat{\cal H}^{r}_{\m\n\r}=
\hat{H}^r_{\m\n\r} -\frac{i}{8}v^r (\bar{\chi}^m \g_{\m\n\r}\chi^m )
\ee satisfies the (anti)self-duality conditions
\be G_{rs} \hat{\cal H}^{s}_{\m\n\r}=\frac{1}{6e}\e_{\m\n\r\a\b\g}
\hat{\cal H}^{\a\b\g}_r \quad. \label{finalselfdual}
\ee
We have also identified the complete supersymmetry transformations, 
that we collect here for convenience:
\beq 
& & \delta e_\m{}^a =-i(\bar{\e} \g^a \Psi_\m ) \quad,\nonumber\\ & & \delta
B^r_{\m\n} =i v^r (\bar{\Psi}_{[\m} \g_{\n]} \e )+ \frac{1}{2} x^{mr} (\bar{\chi}^m
\g_{\m\n} \e ) \quad,
\nonumber\\ & & \delta v_r = x^m_r (\bar{\chi}^m \e )\quad,\nonumber\\ 
 & & \delta \Psi_\m
=\hat{D}_\m \e +\frac{1}{4} v_r \hat{H}^r_{\m\n\r}
\g^{\n\r}\e -\frac{3i}{8} \g_\m \chi^n (\bar{\e} \chi^n ) -\frac{i}{8} \g^\n \chi^n
(\bar{\e}
\g_{\m\n} \chi^n )+\frac{i}{16}
\g_{\m\n\r} \chi^n (\bar{\e} \g^{\n\r} \chi^n ) \quad ,\nonumber\\ & & \delta \chi^m =
\frac{i}{2} x^m_r (\hat{\de_\a v^r} ) \g^\a \e +
\frac{i}{12} x^m_r \hat{H}^r_{\a\b\g} \g^{\a\b\g}\e \quad .
\eeq 

\subsection{Bosonic Equations of Motion and Supersymmetry}

In order to  obtain the bosonic equations, it is convenient to associate the  fermionic
equations to the Lagrangian 
\beq
 e^{-1} {\cal{L}}_{fer} & & = -\frac{i}{2}\bar{\Psi}_\m \g^{\m\n\r} D_\n [\frac{1}{2}(\w
+\hat{\w} )]
\Psi_\r -\frac{i}{8}v_r [H+\hat{H}]^{r \m\n\r}(\bar{\Psi}_\m \g_\n \Psi_\r)
\nonumber \\ & & +\frac{i}{48} v_r [H+\hat{H} ]^r_{\a\b\g} (\bar{\Psi}_\m
\g^{\m\n\a\b\g}\Psi_\n )+\frac{i}{2} \bar{\chi}^m \g^\m D_\m (\hat{\w})
\chi^m \nonumber \\ & & -\frac{i}{24}v_r \hat{H}^r_{\m\n\r} (\bar{\chi}^m \g^{\m\n\r}
\chi^m ) +\frac{1}{4}x^m_r [\de_\n v^r +\hat{\de_\n v^r} ](\bar{\Psi}_\m \g^\n
\g^\m \chi^m) \nonumber\\ & & -\frac{1}{8} x^m_r [H+\hat{H}]^{r \m\n\r} ( \bar{\Psi}_\m
\g_{\n\r}
\chi^m )+\frac{1}{24}x^m_r [H+\hat{H}]^{r \m\n\r} (\bar{\Psi}^\a \g_{\a\m\n\r}
\chi^m ) \nonumber\\ & & +\frac{1}{8}(\bar{\chi}^m \g^{\m\n\r} \chi^m )(\bar{\Psi}_\m
\g_\n \Psi_\r )-\frac{1}{8}(\bar{\chi}^m \g^\a \chi^n )(\bar{\chi}^m \g_\a \chi^n )\quad
,
\label{fermilag}
\eeq where, in the 1.5 order formalism,  the spin connection
\beq & & \w_{\m\n\r} =\w^0_{\m\n\r} -\frac{i}{2}(\bar{\Psi}_\m \g_\n \Psi_\r
+\bar{\Psi}_\n \g_\r
\Psi_\m+\bar{\Psi}_\n \g_\m \Psi_\r ) \nonumber\\ & &  -\frac{i}{4}(\bar{\Psi}^\a 
\g_{\m\n\r\a\b} \Psi^\b )-\frac{i}{4} (\bar{\chi}^m \g_{\m\n\r} \chi^m )
\label{1.5}
\eeq satisfies its equation of motion, and is thus kept fixed in all variations.

In order to  derive the bosonic equations, one can add to (\ref{fermilag})
\be e^{-1}{\cal{L}}_{bose}=-\frac{1}{4}R +\frac{1}{12}G_{rs} H^{r \m\n\r} H^s_{\m\n\r}
-\frac{1}{4} \de_\m v^r \de^\m v_r \quad . \label{boselag}
\ee One can then obtain from ${\cal{L}}_{fer}+{\cal{L}}_{bose}$  the equations for  the
vielbein and the scalars, with the prescription that the  
(anti)self-duality conditions be used
only after varying.  Actually, ignoring momentarily  eq. (\ref{finalselfdual}) and
varying
${\cal{L}}_{fer}+{\cal{L}}_{bose}$  with respect to the antisymmetric tensor
$B^r_{\m\n}$ yields  the second-order tensor equation, the divergence of eq.
(\ref{finalselfdual}),
\beq & & D_\m (G_{rs} \hat{H}^{s \m\n\r}) =\frac{1}{2} D_\m [ x^m_r (\bar{\chi}^m
\g^{\m\n\r\a}
\Psi_\a )] \nonumber\\ & & - \frac{i}{4}D_\m [v_r (\bar{\Psi}_\a \g^{\a\b\m\n\r}\Psi_\b
)]+
\frac{i}{4} D_\m [v_r (\bar{\chi}^m \g^{\m\n\r} \chi^m )] \quad .
\eeq In a similar fashion, the scalar equation is
\beq  & & x^m_r [\frac{1}{2}D_\m (\de^\m v^r ) +\frac{1}{3} v_s H^{r \m\n\r} H^s_{\m\n\r}
-\frac{i}{4} H^{r \m\n\r}(\bar{\Psi}_\m \g_\n \Psi_\r ) 
 +\frac{i}{24}H^r_{\a\b\g} (\bar{\Psi}_\m \g^{\m\n\a\b\g}\Psi_\n )
\nonumber\\ & & -\frac{i}{24} H^r_{\m\n\r} (\bar{\chi}^n \g^{\m\n\r}\chi^n )  -
\frac{1}{2}D_\n (x^n_r(\bar{\Psi}_\m \g^\n \g^\m \chi^n ))]
\nonumber\\  & & + v_r [ -\frac{1}{4} H^{r \m\n\r} (\bar{\Psi}_\m \g_{\n\r} \chi^m ) +
\frac{1}{12} H^{r \m\n\r} (\bar{\Psi}^\a \g_{\a\m\n\r} \chi^m ) ] = 0
\label{scalarcomplete}
\quad ,
\eeq while the Einstein equation is
\beq & & \frac{1}{2}e_{\b a} [R^{\a\b} -\frac{1}{2}g^{\a\b} R -G_{rs}H^{r \a\n\r} H^{s
\b}{}_{\n\r}  +\frac{1}{6}g^{\a\b}G_{rs} H^r_{\m\n\r}H^{s \m\n\r} 
\nonumber\\ & & +\de^\a v^r
\de^\b v_r -\frac{1}{2}g^{\a\b} \de_\m v^r \de^\m v_r ] 
 -\frac{i}{2}e^\a{}_a (\bar{\Psi}_\m \g^{\m\n\r} \hat{D}_\n \Psi_\r )+
\frac{i}{2} (\bar{\Psi}_a \g^{\a\n\r} \hat{D}_\n \Psi_\r )\nonumber\\ & & +\frac{i}{2}
(\bar{\Psi}_\m \g^{\m\a\r} \hat{D}_a \Psi_\r )+
\frac{i}{2} (\bar{\Psi}_\m \g^{\m\n\a} \hat{D}_\n \Psi_a  )-
\frac{i}{4} e^\a{}_a v_r \hat{H}^r_{\m\n\r} (\bar{\Psi}^\m \g^\n \Psi^\r) +\frac{i}{4}v_r
\hat{H}^r_{\m a \r} (\bar{\Psi}^\m \g^\a \Psi^\r ) \nonumber\\ & & +\frac{i}{2} v_r
\hat{H}^{r
\a\n\r} (\bar{\Psi}_a \g_\n \Psi_\r )+
\frac{i}{2}v_r \hat{H}^r_{a \n\r} (\bar{\Psi}^\a \g^\n \Psi^\r )   +\frac{i}{24}
e^\a{}_a v_r
\hat{H}^r_{\b\g\delta} (\bar{\Psi}_\m 
\g^{\m\n\b\g\delta} \Psi_\n )\nonumber\\ & & - \frac{i}{12}v_r \hat{H}^r_{\b\g\delta}
(\bar{\Psi}_a \g^{\a\n\b\g\delta} \Psi_\n )   -\frac{i}{8}v_r \hat{H}^r_{a \b\g}
(\bar{\Psi}_\m
\g^{\m\n\a\b\g} \Psi_\n ) +\frac{i}{2}e^\a{}_a (\bar{\chi}^m \g^\m \hat{D}_\m \chi^m
)\nonumber\\ & & -\frac{i}{2}( \bar{\chi}^m \g^\a \hat{D}_a \chi^m )-\frac{i}{24}
e^\a{}_a v_r
\hat{H}^r_{\m\n\r} (\bar{\chi}^m \g^{\m\n\r} \chi^m )   +\frac{i}{8} v_r \hat{H}^r_{a
\n\r} (\bar{\chi}^m \g^{\a\n\r} \chi^m )
\nonumber\\ & & +
\frac{1}{2} e^\a{}_a x^m_r (\hat{\de_\n v^r} )(\bar{\Psi}_\m \g^\n \g^\m \chi^m ) 
-\frac{1}{2}x^m_r (\hat{\de_a v^r })(\bar{\Psi}_\m \g^\a \g^\m \chi^m ) -\frac{1}{2}x^m_r
(\hat{\de_\n v^r })(\bar{\Psi}_a \g^\n \g^\a \chi^m )
\nonumber\\ & & -\frac{1}{4}e^\a{}_a x^m_r \hat{H}^r_{\m\n\r} (\bar{\Psi}^\m \g^{\n\r} 
\chi^m )+\frac{1}{2} x^m_r \hat{H}^r_{\m a\r}(\bar{\Psi}^\m \g^{\a\r} \chi^m )
 +\frac{1}{4}x^m_r \hat{H}^{r \a}{}_{\n\r}(\bar{\Psi}_a \g^{\n\r} \chi^m )
\nonumber\\ & & + \frac{1}{4}x^m_r \hat{H}^r_{a \n\r}(\bar{\Psi}^\a \g^{\n\r} \chi^m )
 +\frac{1}{12}e^\a{}_a x^m_r \hat{H}^r_{\m\n\r}(\bar{\Psi}_\sigma
\g^{\sigma \m\n\r}\chi^m )-\frac{1}{12}x^m_r \hat{H}^r_{\m\n\r} (\bar{\Psi}_a
\g^{\a\m\n\r}\chi^m ) \nonumber\\ & & -\frac{1}{4}x^m_r \hat{H}^r_{a
\n\r}(\bar{\Psi}_\sigma \g^{\sigma \a\n\r}
\chi^m ) + (fermi)^4 =0 \quad . \label{einsteincomplete}
\eeq For the sake of brevity, a number of quartic fermionic couplings, fully determined
by the lagrangian of eqs. (\ref{fermilag}) and (\ref{boselag}),  are not written
explicitly. It then takes a direct, if somewhat tedious, calculation to prove local
supersymmetry, showing that
\be 
\delta F \frac{\delta {\cal{L}}}{\delta F}+\delta B 
\frac{\delta {\cal{L}}}{\delta B}=0
\quad,\label{susyproof}
\ee where $F$ and $B$ denote collectively the fermi  and  bose fields 
aside from the
antisymmetric tensors. We would like to stress that the equations for the fermi
fields defined from the gauge algebra differ from the lagrangian equations
by overall factors that may be simply identified.

\section{Inclusion of Vector Multiplets} 

A $(1,0)$ Yang-Mills  multiplet in six
dimensions comprises  gauge vectors $A_\m$ and pairs of left-handed spinors $\l^a$
satisfying a  symplectic Majorana condition, all in the adjoint  representation of
the gauge group. In this Section we write the complete field  equations for $N=1$
supergravity coupled to $n$  tensor multiplets and to  vector multiplets,
extending the results of \cite{as,fms}.  This setting plays a crucial role in
six-dimensional perturbative type-I vacua, that naturally include a number of tensor
multiplets \cite{bs}, and more generally in the context of string dualities relating
these to non-perturbative vacua of other strings and to $M$ theory \cite{dmw}.  In all
these cases, the anomaly polynomial comprises in principle an irreducible part, that in
perturbative  type-I vacua is removed by tadpole conditions, and a residual reducible 
part of the form
\be I_8 = - \sum_{x,y} \ c^r_x \ c^s_y \ \eta_{rs} \ {\rm tr}_x F^2 \  {\rm tr}_y F^2
\quad ,
\label{residualanomaly}
\ee with the $c$'s a collection of constants and $\eta$ the Minkowski metric for
$SO(1,n)$.  In general, this residual anomaly should also include gravitational and mixed
contributions, but we leave them aside, since they would contribute higher-derivative
couplings not part of the low-energy effective supergravity.

The antisymmetric tensors are not inert under vector gauge transformations, as demanded
by the Chern-Simons couplings
\be H^r = dB^r - c^{rz} \w_z\quad , \label{newtensorfieldstrength}
\ee where the index $z$ runs over the various factors of the gauge  group.  Gauge
invariance of
$H^r$ indeed requires that $B^r_{\m\n}$ transform under vector gauge transformations
according to
\be
\delta B^r = c^{rz} tr_z (\L dA) \quad .\label{new3form}
\ee To lowest order, the (anti)self-duality conditions (\ref{selfdual}) are not
affected, while their divergence becomes 
\be D_\m (G_{rs} H^{s \m\n\r} ) =-\frac{1}{4e} \e^{\n\r\a\b\g\delta} c_r^z tr_z (F_{\a\b}
F_{\g\delta}) \quad.\label{eqtensorgs}
\ee In a similar fashion, the fermionic equations become
\beq & & \g^{\m\n\r} D_\n \Psi_\r + v_r H^{r \m\n\r} \g_\n \Psi_\r-
\frac{i}{2} x^m_r H^{ r \m\n\r} \g_{\n\r} \chi^m  
\nonumber\\ & & + \frac{i}{2} x^m_r \de_\n v^r \g^\n \g^\m \chi^m -\frac{1}{\sqrt{2}}
v_r c^{rz} tr_z (F_{\sigma \tau} \g^{\sigma \tau} \g^\m \l )=0 
\eeq for the gravitino,
\beq & & \g^\m D_\m \chi^m -\frac{1}{12} v_r H^{r \m\n\r} \g_{\m\n\r} \chi^m
-\frac{i}{2} x^m_r H^{r \m\n\r} \g_{\m\n} \Psi_\r  \nonumber\\ & & - \frac{i}{2} x^m_r
\de_\n v^r \g^\m \g^\n \Psi_\m -\frac{i}{\sqrt{2}} x^m_r c^{rz} tr_z (F_{\m\n}\g^{\m\n}
\l ) =0 
\eeq for the tensorini and
\beq & & v_r c^{rz}\g^\m D_\m \l +\frac{1}{2} (\de_\m v_r ) c^{rz} \g^\m \l+
\frac{1}{2\sqrt{2}} v_r c^{rz} F_{\a\b} \g^\m \g^{\a\b} \Psi_\m \nonumber\\ & & +
\frac{i}{2\sqrt{2}}x^m_r c^{rz} F_{\m\n} \g^{\m\n} \chi^m -
\frac{1}{12} c^{rz} H_{r \m\n\r} \g^{\m\n\r} \l =0 
\eeq for the gaugini. The  supersymmetry transformations of the vector multiplet are
\beq & & \delta A_\m =-\frac{i}{\sqrt{2}} (\bar{\e} \g_\m \l ) \quad ,\nonumber\\ & &
\delta \l =-\frac{1}{2\sqrt{2}} F_{\m\n} \g^{\m\n} \e \quad ,
\eeq while the tensor transformation becomes
\be
\delta B^r_{\m\n} =i v^r (\bar{\Psi}_{[\m} \g_{\n]} \e )+\frac{1}{2} x^{mr} (\bar{\chi}^m
\g_{\m\n} \e )-2 c^{rz} tr_z (A_{[\m} \delta A_{\n]})
\quad . 
\ee The other transformations are not modified, 
aside from the change induced  by
(\ref{newtensorfieldstrength}) in the definition of $H^r$. 
Varying the fermi fields  in the fermionic
equations then gives the bosonic equations
\be x^m_r D_\m (\de^\m v^r ) +\frac{2}{3} x^m_r v_s H^r_{\a\b\g}  H^{s \a\b\g} -x^m_r
c^{rz} tr_z (F_{\a\b} F^{\a\b})=0 
\ee for the scalar,
\beq & & R_{\m\n} -\frac{1}{2} g_{\m\n} R + \de_\m v^r \de_\n v_r -
\frac{1}{2} g_{\m\n} \de_\a v^r \de^\a v_r - G_{rs} H^r_{\m\a\b} H^s{}_\n{}^{\a\b}
\nonumber\\ & & + 4 v_r c^{rz} tr_z (F_{\a\m} F^\a{}_\n -\frac{1}{4} g_{\m\n}
F_{\a\b}F^{\a\b})=0
\eeq for the metric, and
\be D_\m (v_r c^{rz} F^{\m\n} ) -c^{rz} G_{rs} H^{s \n\r \sigma} F_{\r\sigma}=0
\label{vectoreq}
\ee for the vectors. The commutator of two supersymmetry transformations now includes a
gauge  transformation of parameter
\be
\L = \xi^\a A_\a \quad.
\ee

The novelty here is the non-vanishing divergence of eq. (\ref{vectoreq}) 
\be D_\m J^\m = -\frac{1}{2e} \e^{\m\n\a\b\g\delta} c^{rz}c_r^{z^\prime} F_{\m\n}
tr_{z^\prime}(F_{\a\b} F_{\g\delta} ) \quad,\label{covanomaly}
\ee that reflects the presence of the residual gauge anomaly \cite{fms}.  In
particular,  eq. (\ref{covanomaly}) gives the  covariant anomaly.  Leaving aside
momentarily the (anti)self-duality conditions, one  might expect to derive eq.
(\ref{vectoreq}) from
\be e^{-1}{\cal{L}} = -\frac{1}{2} v_r c^{rz} tr_z F_{\m\n} F^{\m\n} +\frac{1}{12}
 G_{rs}H^{r\m\n\r}H^s_{\m\n\r} \quad,
\ee but this is actually not the case. In fact, eq. (\ref{vectoreq}) is not integrable,
while the inclusion of a Wess-Zumino term
\be e^{-1}{\cal{L}} = -\frac{1}{2} v_r c^{rz} tr_z F_{\m\n} F^{\m\n} +\frac{1}{12}
 G_{rs}H^{r\m\n\r}H^s_{\m\n\r} -\frac{1}{8e } \e^{\m\n\a\b\g\delta} c_r^z B^r_{\m\n}
Tr_z (F_{\a\b} F_{\g\delta}) \quad, \label{greenschwarz}
\ee turns the vector equation into
\beq & & D_\m (v_r c^{rz} F^{\m\n} ) -  G_{rs} H^{s \n\r\sigma} c^{rz} F_{\r\sigma}  -
\frac{1}{8e} \e^{\n\r\a\b\g\delta} c_r^z A_\r c^{rz^\prime}  tr_{z^\prime}
(F_{\a\b}F_{\g\delta})
\nonumber\\ & & -\frac{1}{12e}\e^{\n\r\a\b\g\delta} c_r^z F_{\r\a} c^{rz^\prime}
\w^{z^\prime}_{\b\g\delta} = 0 \quad, \label{newvectoreq}
\eeq and now the divergence of the gauge current is the consistent anomaly \cite{fms}
\be {\cal{A}}_\L =- \frac{1}{4} \e^{\m\n\a\b\g\delta} c_r^z c^{rz^\prime} tr_z (\L
\de_\m A_\n ) tr_{z^\prime} (F_{\a\b} F_{\g\delta} )\quad .
\label{consanomaly}
\ee As an aside, one can observe that, ignoring the (anti)self-duality conditions, eq.
(\ref{greenschwarz}) yields the second-order tensor equations (\ref{eqtensorgs}) when
varied with respect to  the antisymmetric fields.

The Wess-Zumino consistency condition \cite{wz}
\be
\delta_{\L} {\cal{A}}_{\e} =\delta_\e {\cal{A}}_\L
\ee now implies the presence of a supersymmetry anomaly of the form
\be {\cal{A}}_\e =-\frac{1}{4} \e^{\m\n\a\b\g\delta} c_r^z c^{r z^\prime} tr_z (
\delta_\e A_\m A_\n ) tr_{z^\prime} (F_{\a\b} F_{\g\delta})  -\frac{1}{6} \e^{\m\n\a\b\g\delta}
c_r^z c^{r z^\prime} tr_z (
\delta_\e A_\m F_{\n\a} ) \w^{z^\prime}_{\b\g\delta} \ ,
\label{susyanomaly}
\ee and indeed the  supersymmetry variation of the lagrangian is exactly eq. 
(\ref{susyanomaly}). Moreover, the divergence of the gravitino field equation,
proportional to eq. (\ref{susyanomaly}), reflects the presence of the induced
supersymmetry anomaly. We shall now complete this construction to all orders  in the
fermi fields.

\subsection{Complete Supersymmetry Algebra} 
Let us begin by noting that the supercovariant  Yang-Mills field strength is
\be
\hat{F}_{\m\n} = F_{\m\n} +\frac{i}{\sqrt{2}}(\bar{\l} \g_\m \Psi_\n ) -
\frac{i}{\sqrt{2}}(\bar{\l} \g_\n \Psi_\m ) \quad ,
\ee while the other supercovariant fields are not modified.  The supersymmetry
transformations 
\beq 
& & \delta e_\m{}^a =-i(\bar{\e} \g^a \Psi_\m ) \quad,\nonumber\\ & & \delta
B^r_{\m\n} =i v^r (\bar{\Psi}_{[\m} \g_{\n]} \e )+ \frac{1}{2} x^{mr} (\bar{\chi}^m
\g_{\m\n} \e )-2c^{rz} tr_z (A_{[\m}\delta A_{\n]}) \quad,
\nonumber\\ & & \delta v_r = x^m_r (\bar{\chi}^m \e )\quad,\nonumber\\ & & \delta A_\m =
-\frac{i}{\sqrt{2}} (\bar{\e} \g_\m \l ) \quad ,\nonumber\\ & & \delta \Psi_\m
=\hat{D}_\m \e +\frac{1}{4} v_r \hat{H}^r_{\m\n\r}
\g^{\n\r}\e -\frac{3i}{8} \g_\m \chi^n (\bar{\e} \chi^n ) -\frac{i}{8} \g^\n \chi^n
(\bar{\e}
\g_{\m\n} \chi^n )+\frac{i}{16}
\g_{\m\n\r} \chi^n (\bar{\e} \g^{\n\r} \chi^n ) \quad ,\nonumber\\ & & \delta \chi^m =
\frac{i}{2} x^m_r (\hat{\de_\a v^r} ) \g^\a \e +
\frac{i}{12} x^m_r \hat{H}^r_{\a\b\g} \g^{\a\b\g}\e \quad ,\nonumber\\ & & \delta \l
=-\frac{1}{2\sqrt{2}}\hat{F}_{\m\n} \g^{\m\n} \e \quad,
\eeq 
could in principle include additional terms proportional to $\l^2$. To be precise,
one could add to $\delta \Psi$ a term proportional to
$v_r c^{rz} tr_z (\l^2 \e)$, and to $\delta \chi$ a term proportional to $x^m_r c^{rz}
tr_z (\l^2
\e )$. Moreover, the (anti)self-duality conditions could be modified by a self-dual term
of the form $c^{rz} tr_z (\bar{\l} \g_{\m\n\r} \l )$.

Let us proceed to study the supersymmetry algebra completely. On the  scalar, the
vielbein and the gauge field, the algebra closes with no subtleties, while additional
information  comes from  the algebra on the tensor fields. Using the (anti)self-duality
conditions satisfied by the 3-forms  in eq. (\ref{newtensor}), one can show that the
algebra on $B^r$ closes up to the extra terms
\be [\delta_1 , \delta_2 ]_{extra} B^r_{\m\n}=c^{rz}tr_z [(\bar{\e}_1 \g_\m \l )
(\bar{\e}_2 \g_\n
\l )-(\bar{\e}_1 \g_\n \l )(\bar{\e}_2 \g_\m \l )] \quad .
\ee These can be  canceled modifying the transformations of the gravitino and of the
tensorini according to
\beq & & \delta^{\prime} \Psi_\m =i v_r c^{rz}\lbrace a \ tr_z [\l (\bar{\e} \g_\m \l
)]+  b \ tr_z [\g_{\m\n} \l (\bar{\e} \g^\n \l )]+ c \ tr_z [\g^{\n\r} \l (\bar{\e}
\g_{\m\n\r} \l )] \rbrace
\quad, 
\nonumber\\ &  & \delta^{\prime}\chi^m = d \ x^m_r c^{rz}tr_z [ \g_\a \l (\bar{\e}
\g^\a \l )] \quad,
\eeq and  requiring that the modified 3-form
\be
\hat{\cal H}^{r}_{\m\n\r} =\hat{H}^{r}_{\m\n\r} -\frac{i}{8} v^r (\bar{\chi}^m
\g_{\m\n\r}\chi^m ) +i\a \ c^{rz} tr_z (\bar{\l} \g_{\m\n\r} \l )
\ee satisfy the (anti)self-duality conditions
\be G_{rs} \hat{\cal H}^{s}_{\m\n\r} =\frac{1}{6e} \e_{\m\n\r\a\b\g}
\hat{\cal H}_r^{\a\b\g}\quad.\label{finalselfdual2}
\ee It should be appreciated that this change in the definition of the field  strengths
only affects the antiself-duality conditions, since
$(\bar{\l}\g_{\a\b\g}\l )$ is self-dual.

Requiring closure of the algebra on $B^r$ then implies the conditions
\be
\a =\frac{1}{4} \quad, 
\qquad d=\frac{1}{2} \quad ,
\qquad a+ b = -1 \quad ,
\qquad b+2c =0 \quad , \label{parameters}
\ee and only one of the parameters is still undetermined. These terms have no effect for
the scalars and the vectors, while the commutator on $e_\m{}^a$ shows that the local
Lorentz parameter is modified by the addition of
\be
\Delta^{\prime} {\W}^{ab} =v_r c^{rz}  tr_z [(\bar{\e}_1 \g^a \l )(\bar{\e}_2 \g^b \l )-
(\bar{\e}_2 \g^a \l )(\bar{\e}_1 \g^b \l ) ] \quad . \label{locallorentz}
\ee

Turning to the fermi fields,    the commutator on the tensorini $\chi^m$ involves
techniques already met in  the case with tensor multiplets only, and fixes the last free
parameter  in eqs. (\ref{parameters}), so that
\be a=-\frac{9}{8}\quad,\qquad b =\frac{1}{8} \quad, \qquad  c =-\frac{1}{16} \quad.
\ee It closes on the field equation
\beq & & \g^\m \hat{D}_\m \chi^m -\frac{1}{12}v_r \hat{H}^r_{\m\n\r} \g^{\m\n\r}
\chi^m -\frac{i}{12} x^m_r \hat{H}^{r \m\n\r} \g_\a \g_{\m\n\r} \Psi^\a
 - \frac{i}{2} x^m_r (\hat{\de_\n v^r}) \g^\m \g^\n \Psi_\m 
\nonumber\\ & & - \frac{i}{\sqrt{2}}x^m_r c^{rz} tr_z (\hat{F}_{\m\n} \g^{\m\n} \l )  
-\frac{1}{2} x^m_r c^{rz} tr_z [ \g^\m \g^\a  \l (\bar{\Psi}_\m \g_\a
\l )]  -\frac{i}{2}\g^\a \chi^n (\bar{\chi}^n \g_\a \chi^m )  
\nonumber\\ & & +\frac{3i}{8} v_r c^{rz} tr_z [(\bar{\chi}^m \g_{\m\n} \l )\g^{\m\n}
\l ] +\frac{i}{4} v_r c^{rz} tr_z [(\bar{\chi}^m \l ) \l ] 
 =0\quad ,\label{tensorinoeq2}
\eeq where all terms containing the gravitino are exactly determined by supercovariance. 
Moreover, the field equation appears in the commutator as in  the theory without gauge
fields:
\be  [\delta_1 , \delta_2 ] \chi^m = \delta_{gct} \chi^m +\delta_{Lorentz}\chi^m
+\delta_{SO(n)}\chi^m +\delta_{susy}\chi^m  + \frac{1}{4} \g^\a \xi_\a [eq.\ \chi^m ]
\quad.
\ee

Using similar techniques, one can compute the commutator on the  gaugini $\l$. Here,
however, the transformation 
\be
\delta \l = -\frac{1}{2\sqrt{2}}\hat{F}_{\m\n}\g^{\m\n}\e\label{gauginovar}
\ee can not produce the terms proportional to $x^m_r$  already present at the lowest
order, and the only way to generate them is to modify eq. (\ref{gauginovar}) by terms of
the form
\be
\frac{x^m_r c^{rz}}{v_s c^{sz}} \ \chi^m \ \l \ \e \quad .
\nonumber
\ee Singular couplings of this type were previously introduced in \cite{ns2}. We
therefore add all possible extra terms, that modulo Fierz identities are
\be
\delta^{\prime} \l =  \frac{x^m_r c^{rz}}{v_s c^{sz}}[a(\bar{\chi}^m \l ) \e+ b (\bar{\chi}^m
\g_{\a\b} \l ) \g^{\a\b}
\e + c (\bar{\chi}^m \e ) \l  +d (\bar{\chi}^m \g_{\a\b} \e ) \g^{\a\b}
\l ]\quad,\label{gauginoextra}
\ee and determine their coefficients from the algebra. Eq. (\ref{gauginoextra}) should
not affect the vector (and, a fortiori, the  tensor) commutator, and thus the
coefficients are  to obey the three equations
\be a -2 c =0\quad , \qquad b=0 \quad ,\qquad c+2d=0\quad .\label{sys1}
\ee The other conditions,
\be a+2b = -\frac{1}{2}\quad ,\qquad c+2d+4b=0\quad ,\qquad 2d +\frac{1}{8}a+
\frac{1}{4}b=\frac{3}{16} \quad , \label{sys2}
\ee are obtained from the algebra on the gaugini, for instance tracking  the terms
generated by eq. (\ref{gauginoextra}) and proportional to  $\de v $. Combining eqs.
(\ref{sys1}) and (\ref{sys2}),  one finally obtains
\be a=-\frac{1}{2} \quad ,\qquad c=-\frac{1}{4} \quad ,\qquad d=\frac{1}{8} \quad .
\ee As was the case for the gravitino already without vector multiplets,  here the
algebra generates the field  equation with a non trivial symplectic structure,
\be
\frac{3}{8} \g^\a \xi_\a [eq. \l^a ] +\frac{1}{96}\g^{\a\b\g}
\sigma^i_b{}^a \xi^i_{\a\b\g} [eq. \l^b ] \quad ,
\ee where $\xi^i_{\a\b\g}$ is defined in eq. (\ref{anomalousxi}).

Eq. (\ref{gauginoextra}) also affects the algebra on  the tensorini, whose field
equation now includes two additional terms, and becomes 
\beq & & \g^\m \hat{D}_\m \chi^m -\frac{1}{12}v_r \hat{H}^r_{\m\n\r} \g^{\m\n\r}
\chi^m -\frac{i}{12} x^m_r \hat{H}^{r \m\n\r} \g_\a \g_{\m\n\r} \Psi^\a
 - \frac{i}{2} x^m_r (\hat{\de_\n v^r}) \g^\m \g^\n \Psi_\m 
\nonumber\\ & & - \frac{i}{\sqrt{2}}x^m_r c^{rz} tr_z (\hat{F}_{\m\n} \g^{\m\n} \l ) 
 -\frac{1}{2} x^m_r c^{rz} tr_z [ \g^\m \g^\a  \l (\bar{\Psi}_\m \g_\a
\l )] -\frac{i}{2}\g^\a \chi^n (\bar{\chi}^n \g_\a \chi^m ) \nonumber\\ & & +
\frac{3i}{8}
v_r c^{rz} tr_z [(\bar{\chi}^m \g_{\m\n} \l )\g^{\m\n}
\l ] +\frac{i}{4} v_r c^{rz} tr_z [(\bar{\chi}^m \l ) \l ] \nonumber\\ & & +
\frac{3i}{2} \frac{x^m_r c^{rz} x^n_s c^{sz}}{v_t c^{tz}} tr_z [(\bar{\chi}^n \l ) \l ]  -
\frac{i}{4}
\frac{x^m_r c^{rz} x^n_s c^{sz}}{v_t c^{tz}} tr_z [(\bar{\chi}^n \g_{\a\b} \l )
\g^{\a\b} \l ]
 =0\quad .\label{tensorinoeq3}
\eeq

In the commutator of two supersymmetry transformations on the gaugini, these terms
complete the algebra and let it close on the field equation, that now includes 
$\chi^2 \l$  terms corresponding to the $\l^2 \chi$ terms in the equation for  the
tensorini. In addition, the $\l^3$ terms comprise two groups:  those proportional to 
$v_r v_s$ and those proportional to $\eta_{rs}$ (recall, from eqs.  (\ref{scalars}),
that 
$x^m_r x^m_s = v_r v_s -\eta_{rs}$). The former generate local Lorentz transformations
according to eq. (\ref{locallorentz}) and the term
\be iv_r v_s c^{rz} c^{s z^\prime} tr_{z^\prime} [(\bar{\l} \g_\a \l^\prime )
\g^\a \l^\prime ]
\ee in the field equation, while the latter are 
\beq [\delta_1 , \delta_2 ]_{extra} \l & & = \frac{c_r^z c^{r z^\prime}}{v_s c^{sz}}
tr_{z^\prime} [-\frac{1}{4}(\bar{\e}_1 \g_\a \l^\prime )(\bar{\e}_2
\g_\b \l^\prime ) \g^{\a\b} \l  +
\frac{1}{4} (\bar{\l} \g_\a \l^\prime ) (\bar{\e}_1 \g^\a \l^\prime ) \e_2 -(1
\leftrightarrow 2)
\nonumber\\ & &  + \frac{1}{16} (\bar{\e}_1 \g^\a \e_2 )(\bar{\l}^\prime
\g_{\a\b\g} \l^\prime  )\g^{\b\g}\l ]\quad . \label{centralcharge1}
\eeq In general, one could allow for a modified field equation including the 
$\l^3$ term
\be -i \a c_r^z c^{r z^\prime} tr_{z^\prime} [(\bar{\l}\g_\a \l^\prime )
\g^\a \l^\prime ] \quad , \label{lambda3}
\ee with $\a$ an arbitrary parameter. Although the choice $\a =1$ could seem  the
preferred one on account of the rigid limit, since the  supersymmetric Yang-Mills theory
in six dimensions does not contain such a $\l^3$ term,  the $(1,0)$ supergravity is
actually consistent for an arbitrary value  of $\a$, with the corresponding residual
terms
\beq
 \delta_{extra(\a)} \l & &\equiv [\delta_1 , \delta_2 ]_{extra(\a)} \l = \frac{c_r^z c^{r
z^\prime}}{v_s c^{sz}} tr_{z^\prime} [-\frac{1}{4}(\bar{\e}_1 \g_\a \l^\prime
)(\bar{\e}_2
\g_\b \l^\prime ) \g^{\a\b} \l \nonumber\\  & &-\frac{\a}{2} (\bar{\l} \g_\a \l^\prime
)(\bar{\e}_1 \g_\b \l^\prime )
\g^{\a\b} \e_2  +\frac{\a}{16}(\bar{\l}\g_{\a\b\g}\l^\prime )(\bar{\e}_1 \g^\g 
\l^\prime ) \g^{\a\b} \e_2 \nonumber\\  & &+\frac{\a}{16} (\bar{\l} \g_\g \l^\prime
)(\bar{\e}_1
\g^{\a\b\g} \l^\prime ) \g_{\a\b} \e_2   + \frac{1-\a}{4} (\bar{\l} \g_\a \l^\prime )
(\bar{\e}_1
\g^\a \l^\prime ) \e_2 -(1 \leftrightarrow 2)  
\nonumber\\ & & + \frac{1-\a}{16} (\bar{\e}_1 \g^\a \e_2 )(\bar{\l}^\prime
\g_{\a\b\g} \l^\prime  )\g^{\b\g}\l ] \label{centralcharge}
\eeq in the commutator of two supersymmetry transformations on the gaugini.   It should
be appreciated that no choice of $\a$ can eliminate all these terms, that  play the role
of a central charge felt only by the gaugini. The Jacobi  identity for this charge is
properly satisfied for any value of
$\a$, and thus we are effectively discovering a 2-cocycle in our problem. It has long
been known that, in general, anomalies in current conservations are accompanied by
related anomalies in current commutators \cite{anom}, but it is amusing to see how this
``classically anomalous'' model displays all these intricacies.

The complete algebra 
\beq [ \delta_1, \delta_2 ] \l^a & & = \delta_{gct} \l^a + \delta_{Lorentz} \l^a +
\delta_{susy}
\l^a + \delta_{gauge} \l^a + \delta_{extra(\a)} \l^a \nonumber \\ & & + \frac{3}{8}
\g^\a \xi_\a [eq. \l^a ]_{(\a )} +\frac{1}{96}\g^{\a\b\g}
\sigma^i_b{}^a \xi^i_{\a\b\g} [eq. \l^b ]_{(\a )} 
\eeq determines the complete field equation of the gaugini
\beq & & v_r c^{rz} \g^\m \hat{D}_\m \l +\frac{1}{2}(\hat{\de_\m v_r})c^{rz}
\g^\m \l + \frac{1}{2\sqrt{2}}v_r c^{rz} \hat{F}_{\a\b} \g^\m \g^{\a\b} \Psi_\m +
\frac{i}{2\sqrt{2}}x^m_r c^{rz} \hat{F}_{\a\b} \g^{\a\b}\chi^m \nonumber\\ & &
+\frac{1}{12} x^m_r c^{rz} x^m_s \hat{H}^s_{\m\n\r} \g^{\m\n\r}\l +
\frac{1}{2}x^m_r c^{rz} (\bar{\chi}^m \l ) \g^\m \Psi_\m 
 +\frac{1}{4}x^m_r c^{rz} (\bar{\chi}^m \Psi_\m )\g^\m \l \nonumber\\ & & - \frac{1}{8}
x^m_r c^{rz} (\bar{\chi}^m \g_{\a\b} \Psi_\m ) \g^{\m\a\b} \l 
 - \frac{1}{4}x^m_r c^{rz} (\bar{\chi}^m \g_{\m\a}\Psi^\m )\g^\a \l+
\frac{i}{8}v_r c^{rz} (\bar{\l} \chi^m )\chi^m \nonumber\\ & & +\frac{3i}{16} v_r c^{rz}
(\bar{\l} \g_{\a\b} \chi^m ) \g^{\a\b} \chi^m +\frac{3i}{4} \frac{x^m_r c^{rz} x^n_s
c^{sz}}{v_t c^{tz}} (\bar{\l} \chi^m )
\chi^n  -\frac{i}{8} \frac{x^m_r c^{rz} x^n_s c^{sz}}{v_t c^{tz}} (\bar{\l} 
\g_{\a\b} \chi^m ) \g^{\a\b} \chi^n \nonumber \\ & & + iv_r v_s c^{rz} c^{s z^\prime}
tr_{z^\prime} [(\bar{\l} \g_\a \l^\prime )
\g^\a \l^\prime ] -i \a c_r^z c^{r z^\prime} tr_{z^\prime} [(\bar{\l}\g_\a \l^\prime )
\g^\a \l^\prime ] = 0  
\label{gauginoeq}
\eeq where, again, all terms containing the gravitino are fixed by supercovariance,
while the
$\chi^2 \l$ terms are precisely as demanded by the 
$\l^2 \chi$ terms in the field equations of the tensorini. At last, one can study the
algebra on the gravitino, thus obtaining the  field equation
\beq & & \g^{\m\n\r} \hat{D}_\n \Psi_\r +\frac{1}{4} v_r \hat{H}^r_{\n\a\b}
\g^{\m\n\r}\g^{\a\b} \Psi_\r -\frac{i}{12}x^m_r \hat{H}^{r \a\b\g}\g_{\a\b\g} \g^\m
\chi^m +\frac{i}{2} x^m_r (\hat{\de_\n v^r })\g^\n \g^\m \chi^m \nonumber\\ & &
+\frac{3i}{2}
\g^{\m\a}\chi^m (\bar{\chi}^m \Psi_\a )-\frac{i}{4}
\g^{\m\a} \chi^m (\bar{\chi}^m \g_{\a\b} \Psi^\b ) 
 +\frac{i}{4} \g_{\a\b} \chi^m (\bar{\chi}^m \g^{\m\a} \Psi^\b )-\frac{i}{2}
\chi^m (\bar{\chi}^m \g^{\m\a} \Psi_\a ) \nonumber\\ & & +v_r c^{rz} tr_z
[-\frac{1}{\sqrt{2}}\g^{\a\b}\g^\m \l \hat{F}_{\a\b}  + \frac{3i}{4} \g^{\m\n\r}\l
(\bar{\Psi}_\n
\g_\n \l )  -\frac{i}{2} \g^\m \l (\bar{\Psi}_\n \g^\n \l ) +\frac{i}{2}\g^\n \l 
(\bar{\Psi}_\n
\g^\m \l ) \nonumber\\ & & + \frac{i}{4}\g_\r \l (\bar{\Psi}_\n \g^{\m\n\r}\l )]
 +\frac{1}{2} x^m_r c^{rz} tr_z [\g_\a \l (\bar{\chi}^m \g^\a \g^\m \l )]=0
\quad,
\eeq that enters the supersymmetry algebra as in eq. (\ref{gravitinoalg}). Once more,
all terms containing the gravitino are fixed by supercovariance, while the other $\l^2
\chi$ terms are precisely as demanded by the 
$\l^2 \Psi$ terms in the tensorino equation and by the $\l \Psi \chi$ terms in the
equations of the gaugini. 

Summarizing, from the algebra we have obtained the complete fermionic equations of
$(1,0)$ six-dimensional supergravity coupled to vector and tensor multiplets. 
In addition, the modified 3-form
\be
\hat{\cal H}^{r}_{\m\n\r}=
\hat{H}^r_{\m\n\r} -\frac{i}{8}v^r (\bar{\chi}^m \g_{\m\n\r}\chi^m )
+ \frac{i}{4} c^{rz} tr_z( \bar{\l} \g_{\m\n\r} \l )
\ee 
satisfies the (anti)self-duality conditions
\be G_{rs} \hat{\cal H}^{s}_{\m\n\r}=\frac{1}{6e}\e_{\m\n\r\a\b\g}
\hat{\cal H}^{\a\b\g}_r \quad. \label{finalselfdual3}
\ee
We have also identified the complete supersymmetry transformations, 
that we collect here for convenience:
\beq
& & \delta e_\m{}^a =-i(\bar{\e} \g^a \Psi_\m ) \quad,\nonumber\\ 
& & \delta
B^r_{\m\n} =i v^r (\bar{\Psi}_{[\m} \g_{\n]} \e )+ \frac{1}{2} x^{mr} (\bar{\chi}^m
\g_{\m\n} \e )-2c^{rz} tr_z (A_{[\m}\delta A_{\n]}) \quad,
\nonumber\\ 
& & \delta v_r = x^m_r (\bar{\chi}^m \e )\quad,\nonumber\\ 
& & \delta A_\m =
-\frac{i}{\sqrt{2}} (\bar{\e} \g_\m \l ) \quad ,\nonumber\\ 
& & \delta \Psi_\m =\hat{D}_\m \e +\frac{1}{4} v_r \hat{H}^r_{\m\n\r}
\g^{\n\r}\e -\frac{3i}{8} \g_\m \chi^n (\bar{\e} \chi^n ) -\frac{i}{8} \g^\n \chi^n
(\bar{\e} \g_{\m\n} \chi^n )+\frac{i}{16} \g_{\m\n\r} \chi^n (\bar{\e} 
\g^{\n\r} \chi^n ) \nonumber \\
& & \quad \quad - \frac{9i}{8} v_r c^{rz} tr_z [\l (\bar{\e} \g_\m \l)] +  
\frac{i}{8} v_r c^{rz} tr_z [\g_{\m\n} \l (\bar{\e} \g^\n \l )]
- \frac{i}{16} v_r c^{rz} tr_z [\g^{\n\r} \l (\bar{\e}
\g_{\m\n\r} \l )] \quad ,\nonumber\\ 
& & \delta \chi^m =
\frac{i}{2} x^m_r (\hat{\de_\a v^r} ) \g^\a \e +
\frac{i}{12} x^m_r \hat{H}^r_{\a\b\g} \g^{\a\b\g}\e +
\frac{1}{2} x^m_r c^{rz} tr_z [ \g_\a \l (\bar{\e} \g^\a \l ) ] \quad ,\nonumber\\ 
& & \delta \l
=-\frac{1}{2\sqrt{2}}\hat{F}_{\m\n} \g^{\m\n} \e 
- \frac{1}{2} \frac{x^m_r c^{rz}}{v_s c^{sz}} (\bar{\chi}^m \l ) \e 
 - \frac{1}{4} \frac{x^m_r c^{rz}}{v_s c^{sz}} (\bar{\chi}^m \e ) \l  
\nonumber \\
& & \quad \quad + \frac{1}{8} \frac{x^m_r c^{rz}}{v_s c^{sz}} (\bar{\chi}^m \g_{\a\b} \e ) \g^{\a\b}
\l \quad .
\eeq 

\subsection{Bosonic Equations of Motion} Proceeding as in Section 2, the bosonic
equations  can be derived from a lagrangian, with the prescription of using the tensor
(anti)self-duality conditions only after varying.  The lagrangian is obtained
supplementing ${\cal{L}}_{fermi}+{\cal{L}}_{bose}$  of eqs. (\ref{fermilag}) and
(\ref{boselag}) with the terms
\beq  & & -\frac{1}{2} v_r c^{rz} tr_z (F_{\m\n} F^{\m\n}) -\frac{1}{8e}
\e^{\m\n\a\b\g\delta} c_r^z B^r_{\m\n} Tr_z (F_{\a\b} F_{\g\delta}) \nonumber \\  & & +
\frac{i}{2\sqrt{2}}  v_r c^{rz} tr_z [(F+\hat{F})_{\sigma \delta}(\bar{\Psi}_\m
\g^{\sigma \delta}\g^\m \l )] 
 +\frac{1}{\sqrt{2}}x^m_r c^{rz} tr_z [(\bar{\chi}^m \g^{\m\n}
\l )\hat{F}_{\m\n} ] \nonumber\\ & & +iv_r c^{rz} tr_z [(\bar{\l} \g^m \hat{D}_\m \l
)+\frac{i}{12} x^m_r x^m_s \hat{H}^r_{\m\n\r} c^{sz} tr_z (\bar{\l}\g^{\m\n\r} \l )
 +\frac{1}{16}v_r c^{rz}tr_z (\bar{\l} \g_{\m\n\r} \l )(\bar{\chi}^m
\g^{\m\n\r} \chi^m ) \nonumber\\ & & - \frac{i}{8}(\bar{\chi}^m \g_{\m\n}\Psi_\r )x^m_r
c^{rz} tr_z (\bar{\l} \g^{\m\n\r} \l ) 
 - \frac{i}{2} x^m_r c^{rz} tr_z [(\bar{\chi}^m \g^\m \g^\a \l ) (\bar{\Psi}_\m \g_\a \l
)]
\nonumber\\ & & - \frac{3}{16}v_r c^{rz}tr_z [(\bar{\chi}^m \g_{\m\n} \l ) (\bar{\chi}^m
\g^{\m\n}
\l )] 
 -\frac{1}{8} v_r c^{rz} tr_z [(\bar{\chi}^m \l )(\bar{\chi}^m \l)
 \nonumber\\ & & -\frac{3}{4}\frac{x^m_r c^{rz}x^n_s c^{sz}}{v_t c^{tz}} tr_z
[(\bar{\chi}^m \l )(\bar{\chi}^n \l )] 
 +\frac{1}{8}\frac{x^m_r c^{rz} x^n_s c^{sz}}{v_t c^{tz}} tr_z [(\bar{\chi}^m \g_{\a\b}
\l )(\bar{\chi}^n \g^{\a\b}\l )]
\nonumber\\ & & +\frac{1}{4} (\bar{\Psi}_\m \g_\n \Psi_\r )(\bar{\l} \g^{\m\n\r}
\l )  -\frac{1}{2} v_r v_s c^{rz}c^{s z^\prime} tr_{z,z^\prime} [(\bar{\l}\g_\a \l^\prime
)(\bar{\l} \g^\a \l^\prime ) ]
 \nonumber\\ & & +\frac{\a}{2}c^{rz} c_r^{z^\prime} tr_{z,z^\prime} [(\bar{\l} \g_\a
\l^\prime )(\bar{\l} \g^\a \l^\prime )] \quad ,
\label{completelag}
\eeq and the 1.5 order formalism  requires that the spin connection $\w_{\m\n\r}$ now
include the additional term
\be
\w^{(\l )}_{\m\n\r} = -\frac{i}{2} v_r c^{rz} tr_z (\bar{\l} \g_{\m\n\r} \l )
\quad .
\ee With the new definition of $\w$, eqs. (\ref{fermilag}), (\ref{boselag}) and 
(\ref{completelag}) then yield the fermi  equations. Moreover,  varying with respect to
$B^r_{\m\n}$ yields the second-order tensor equations, the divergence of the
(anti)self-duality conditions.  The  vector equation is covariant, aside from the
anomalous couplings introduced by the Wess-Zumino term in  eq. (\ref{greenschwarz}). The 
complete residual gauge anomaly is thus given in eq. (\ref{consanomaly}). As we shall
see, it solves the Wess-Zumino consistency conditions even in the presence of
supersymmetry. 

The complete vector field equation is
\beq & & c^{rz} D_\n (v_r F^{\n\m} ) - G_{rs}\hat{H}^{r \m\n\r} c^{sz} F_{\n\r}
-\frac{1}{12e}\e^{\m\n\r\a\b\g} c_r^z c^{r z^\prime} F_{\n\r} \w^{z^\prime}_{\a\b\g}
\nonumber\\ & & -\frac{1}{8e}\e^{\m\n\r\a\b\g} c_r^z c^{r z^\prime} A_\n tr_{z^\prime}
(F_{\r\a} F_{\b\g} )
 -\frac{i}{4}v_r c^{rz} F_{\n\r} (\bar{\Psi}_\a \g^{\a\b\m\n\r} 
\Psi_\b ) \nonumber\\ & & + \frac{i}{4} v_r c^{rz} F_{\n\r} (\bar{\chi}^m \g^{\m\n\r}
\chi^m )
 -\frac{x^m_r c^{rz}}{2} F_{\n\r} (\bar{\Psi}_\a \g^{\a\m\n\r} \chi^m ) 
- \frac{i}{2} x^m_r x^m_s c^{rz} c^{s
z^\prime} F_{\n\r} tr_{z^\prime} (
\bar{\l} \g^{\m\n\r} \l ) \nonumber\\ & & +\frac{i}{\sqrt{2}}  c^{rz} D_\n [v_r (\bar{\Psi}_\a
\g^{\m\n}
\g^\a \l )] +\frac{1}{\sqrt{2}} c^{rz}D_\n  [x^m_r (\bar{\chi}^m \g^{\m\n} \l )] = 0 \quad ,
\label{completevectoreq}
\eeq the complete scalar equation is obtained adding to eq. (\ref{scalarcomplete})  the
terms
\beq  && x^m_r \lbrace \frac{1}{32} c^{rz} tr_z (\bar{\l} \g_{\n\a\b}\l )(\bar{\Psi}_\m
\g^{\m\n\r}\g^{\a\b} \Psi_\r) +\frac{i}{2\sqrt{2}}c^{rz} tr_z[(\bar{\Psi}_\m \g^{\a\b} \g^\m
\l )(F +\hat{F} )_{\a\b} )]  \nonumber\\ && + i c^{rz}tr_z (\bar{\l} \g^\m \hat{D}_\m \l
) -\frac{3}{16} c^{rz} tr_z  [(\bar{\chi}^n \g^{\a\b} \l )(\bar{\chi}^n \g_{\a\b} \l )]
-\frac{1}{8} c^{rz} tr_z  [(\bar{\chi}^n \l )(\bar{\chi}^n \l )]\nonumber\\ &&
+\frac{1}{4} c^{rz} tr_z (\bar{\l} \g^{\m\n\r} 
\l ) (\bar{\Psi}_\m \g_\n \Psi_\r ) - v_s c^{rz} c^{s z^\prime}tr_{z, z^\prime}
[(\bar{\l} \g^\a
\l^\prime )(\bar{\l} \g_\a \l^\prime )]\nonumber\\ && - \frac{1}{8}\frac{x^n_s c^{s z}
x^p_t c^{t z}}{(v\cdot c^z )^2 }c^{rz} tr_z [(\bar{\chi}^n \g^{\a\b} \l )(\bar{\chi}^p
\g_{\a\b} \l )] +\frac{3}{4} \frac{x^n_s c^{s z} x^p_t c^{t z}}{(v\cdot c^z )^2 }c^{rz}
tr_z [(\bar{\chi}^n \l )(\bar{\chi}^p \l )] \rbrace \nonumber\\ && +v_r \lbrace
\frac{1}{\sqrt{2}}c^{rz} tr_z [(\bar{\chi}^m \g^{\a\b} \l )
\hat{F}_{\a\b} ] +\frac{i}{12} x^m_s \hat{H}^{r \m\n\r} c^{sz} tr_z (\bar{\l}
\g_{\m\n\r }  \l )\nonumber\\ &&+ \frac{i}{12} x^m_s \hat{H}^{s \m\n\r} c^{rz} tr_z
(\bar{\l}
\g_{\m\n\r }  \l ) -\frac{i}{8} c^{rz} tr_z (\bar{\l} \g_{\m\n\r} \l ) (\bar{\chi}^m
\g^{\m\n}
\Psi^\r )\nonumber\\ && -\frac{i}{2} c^{rz} tr_z [(\bar{\chi}^m \g^\m \g^\a \l
)(\bar{\Psi}_\m
\g_\a \l )] +\frac{1}{4}\frac{x^n_s c^{rz}c^{sz}}{v_t c^{tz}}tr_z  [(\bar{\chi}^m
\g^{\a\b} \l )(\bar{\chi}^n \g_{\a\b} \l )]\nonumber\\ &&-\frac{3}{2}\frac{x^n_s
c^{rz}c^{sz}}{v_t c^{tz}}tr_z [(\bar{\chi}^m \l ) (\bar{\chi}^n \l )] \rbrace \quad ,
\eeq and the Einstein equation is obtained adding to eq. (\ref{einsteincomplete})  the
terms
\beq && c^{rz} tr_z \lbrace 2e_{\b a} v_r ( F_{\g\a} F^\g{}_\b -\frac{1}{2}g_{\a\b} 
F_{\g\delta} F^{\g\delta} ) + 
\frac{i}{\sqrt{2}} e^\a{}_a v_r (\bar{\Psi}_\m \g^{\b\g} \g^\m \l )F_{\b\g}\nonumber\\ &&
-\frac{2i}{\sqrt{2}} v_r (\bar{\Psi}_\m \g^{\a\g} \g^\m \l ) F_{a \g}-\frac{i}{\sqrt{2}}
v_r (\bar{\Psi}_a \g^{\b\g} \g^\a \l ) F_{\b\g}  +\frac{1}{\sqrt{2}} x^m_r e^\a{}_a
(\bar{\chi}^m
\g^{\b\g} \l ) F_{\b\g}\nonumber\\ && - \sqrt{2} x^m_r (\bar{\chi}^m \g^{\a\g}
\l ) F_{a
\g} +i e^\a{}_a v_r (\bar{\l} \g^\m D_\m \l ) -iv_r (\bar{\l} \g^\a D_a \l )\nonumber\\
&& +\frac{i}{12} e^\a{}_a x^m_r x^m_s H^s_{\m\n\r} (\bar{\l} \g^{\m\n\r} \l )
-\frac{i}{4} x^m_r x^m_s H^s_{a \n\r} (\bar{\l} \g^{\a\n\r} \l )\nonumber\\ &&
-\frac{i}{4} e_{\b a}D_\r [v_r (\bar{\l}\g^{\a\b\r} \l )]+(fermi)^4 \rbrace
\quad .
\eeq We would like to stress that this result is expressed  in terms of the previous
definition of
$\w$, not corrected by bilinears in the gaugini.  Moreover, for the sake of brevity, a
number of quartic fermionic couplings, fully determined by the lagrangian of eqs.
(\ref{fermilag}), (\ref{boselag}) and  (\ref{completelag}), are not written explicitly.
Letting $F$ and $B$ denote all the fermi and bose fields aside from  the antisymmetric 
tensors, the supersymmetry variation of the lagrangian, after using the
(anti)self-duality conditions of eq. (\ref{finalselfdual2}), is
\be
\delta B \frac{\delta {\cal{L}}}{\delta B} +\delta F 
\frac{\delta {\cal{L}}}{\delta F }={\cal{A}}_\e \quad,
\ee where ${\cal{A}}_\e$ is the complete supersymmetry anomaly. Neglecting the last term
in eq. (\ref{completelag}), {\it i.e.} setting $\a$ to zero, 
\beq {\cal{A}}_\e & &=c_r^z c^{r z^\prime} tr_{z, z^\prime} \lbrace -\frac{1}{4}
\e^{\m\n\a\b\g\delta}\delta_\e A_\m A_\n F^\prime_{\a\b} F^\prime_{\g\delta} -\frac{1}{6}
\e^{\m\n\a\b\g\delta} \delta_\e A_\m F_{\n\a} 
\w^\prime_{\b\g\delta} \nonumber\\ & & +\frac{i e}{2} \delta_\e A_\m F_{\n\r}
(\bar{\l}^\prime
\g^{\m\n\r} 
\l^\prime )+\frac{i e}{2} \delta_\e A_\m (\bar{\l} \g^{\m\n\r} \l^\prime )
F^\prime_{\n\r}  + ie\delta_\e A_\m (\bar{\l}\g_\n \l^\prime ) F^{\prime \m\n}
\nonumber\\ & & +
\frac{e}{32} \delta_\e e_\m{}^a (\bar{\l} \g^{\m\a\b} \l )(\bar{\l}^\prime
\g_{a \a\b} \l^\prime )  -\frac{e}{2\sqrt{2}} \delta_\e A_\a (\bar{\l} \g^\a \g^\b \g^\g 
\l^\prime )(\bar{\l}^\prime \g_\b \Psi_\g )  \nonumber\\ & & + \frac{e x^m_s c^{s
z^\prime}}{v_t c^{t z^\prime}} [-\frac{3 i}{2\sqrt{2}}
 \delta_\e A_\a (\bar{\l} \g^\a \l^\prime )(\bar{\l}^\prime \chi^m )
 -\frac{i}{4 \sqrt{2}} \delta_\e A_\a (\bar{\l} \g^{\a\b\g} \l^\prime )(\bar{\l}^\prime
\g_{\b\g}\chi^m ) \nonumber\\ & & - \frac{i}{2\sqrt{2}} \delta_\e A_\a  (\bar{\l} \g_\b
\l^\prime )(\bar{\l}^\prime \g^{\a\b} \chi^m )] \rbrace \quad ,
\label{theanomaly}
\eeq while including the last term in eq. (\ref{completelag}) would give the additional
contribution
\be
\Delta {\cal{A}}_\e = \delta_\e {\cal{L}}_{\l^4} \quad ,\label{extraanomaly}
\ee where
\be {\cal{L}}_{\l^4} =  \frac{e \a}{2} \ c_r^z c^{r z^\prime} tr_{z, z^\prime}
[(\bar{\l} \g^\a
\l^\prime )(\bar{\l}\g_\a \l^\prime ) ] \quad.
\ee 

In verifying the supersymmetry anomaly, the equations for the fermi
fields and for the vector field are presented here must be rescaled by suitable
overall factors that may be simply identified.

We now turn to show that ${\cal{A}}_\e$ satisfies the complete Wess-Zumino
consistency conditions.

\subsection{Wess-Zumino Consistency Conditions} In general, the Wess-Zumino consistency
conditions follow from the requirement that the symmetry algebra be realized on the
effective action. For locally supersymmetric theories this implies
\beq & & \delta_{\L_1} {\cal{A}}_{\L_2} -\delta_{\L_2} {\cal{A}}_{\L_1}=
{\cal{A}}_{[\L_1 ,\L_2 ]}
\quad ,\nonumber \\ & & \delta_\e {\cal{A}}_\L = \delta_\L {\cal{A}}_\e \quad ,\nonumber
\\ & &
\delta_{\e_1} {\cal{A}}_{\e_2} -\delta_{\e_2} {\cal{A}}_{\e_1}= {\cal{A}}_{\tilde{\e}}+
{\cal{A}}_{\tilde{\L}} \quad ,\label{wesszumino}
\eeq where only gauge and supersymmetry anomalies are considered, and where 
$\tilde{\e}$ and $\tilde{\L}$ are the parameters of supersymmetry and gauge
transformations determined by the supersymmetry algebra.

In global supersymmetry the analysis is somewhat simpler, since the r.h.s. of the  last
of eqs. (\ref{wesszumino}) does not contain the (global) supersymmetry anomaly.  Let us
therefore begin by reviewing the case of supersymmetric  Yang-Mills theory in four
dimensions \cite{wzsusy}.  From the 6-form anomaly polynomial
\be I_6 =tr F^3 \quad ,
\ee in the language of forms, one obtains the four-dimensional gauge anomaly
\be {\cal{A}}^{(4)}_\L =tr [\L (dA)^2 +\frac{ig}{2} \ d\L A^3 ] \quad,
\ee and from eqs. (\ref{wesszumino}) one can determine the form of the global
supersymmetry anomaly.  With the classical lagrangian
\be  {\cal{L}}_{SYM} ={\rm tr} \left[ -\frac{1}{2} F_{\m\n} F^{\m\n} + 2 i 
\bar{\l} \g^\m D_\m \l \right]  \quad ,
\ee and $\l$ a right-handed Weyl spinor, the supersymmetry transformations are
\beq & & \delta A_\m =\frac{i}{\sqrt{2}}(\bar{\e} \g_\m \l -\bar{\l}
\g_\m \e )\quad ,\nonumber\\ & & \delta \l =\frac{1}{2\sqrt{2}} F_{\m\n} \g^{\m\n} \e
\quad .
\eeq The second of eqs. (\ref{wesszumino}) (with ${\cal A}_{\tilde{\e}}$ absent in this
global case), then determines the supersymmetry anomaly up to terms cubic in $\l$,
\be {\cal{A}}^{(4)}_\e =tr [\delta_\e A A (dA) +\delta_\e A(dA) A-\frac{3ig}{2}\delta_\e A
A^3 ]\quad ,
\ee and indeed
\be
\delta_{\e_2}{\cal{A}}^{(4)}_{\e_1} -\delta_{\e_1} {\cal{A}}^{(4)}_{\e_2}
={\cal{A}}^{(4)}_{\tilde{\L}}+3tr [\delta_{\e_1} A \delta_{\e_2} A F -
\delta_{\e_2} A \delta_{\e_1} A F ] \quad . \label{susy4dpartial}
\ee In order to compensate the second term in eq. (\ref{susy4dpartial}), one is to add to
${\cal{A}}^{(4)}_\e$ the gauge-invariant term
\be
\Delta {\cal{A}}^{(4)}_\e = -\frac{i}{2}tr [\delta_\e A \bar{\l} \g^{(3)} \l + \bar{\l}
\delta_\e A
\g^{(3)} \l ] \quad ,
\ee so that ${\cal{A}}^{(4)}_\e  +\Delta {\cal{A}}^{(4)}_\e$ is the proper global
supersymmetry anomaly. Although the supersymmetry algebra closes only on the  field
equation of $\l$, in four dimensions a simple dimensional counting shows that eqs.
(\ref{wesszumino}) can not generate a term proportional to $\g^\m D_\m \l$. Therefore,
in this case the Wess-Zumino consistency conditions close  accidentally even off-shell,
as pointed out in \cite{wzsusy}.

The situation is quite different in six dimensions. In this case, in the spirit of the 
previous Section, let us restrict our attention to the 8-form residual anomaly
polynomial 
\be I_8 = - \  c^{rz} c_r^{z^\prime} tr_z (F^2 ) tr_{z^\prime} (F^2 ) \quad ,
\ee where the sums are left implicit, so that the gauge anomaly is
\be {\cal{A}}^{(6)}_{\L} =-c^{rz} c_r^{z^\prime} tr_z (\L dA ) tr_{z^\prime} (F^2 ) \quad .
\ee Then, from the second of eqs. (\ref{wesszumino}),
\be {\cal{A}}^{(6)}_\e =-c^{rz} c_r^{z^\prime } [ tr_z (\delta_\e A A ) tr_{z^\prime} (F^2 )
+2 tr_z (\delta_\e AF )
\w^{z^\prime}_3 ]\quad , \label{d=6globalanomaly}
\ee but there are residual terms, so that
\be (\delta_{\e_1} {\cal{A}}^{(6)}_{\e_2} -\delta_{\e_2}
{\cal{A}}^{(6)}_{\e_1} )_{extra} =-4 c^{rz} c_r^{z^\prime } [ tr_z
(\delta_{\e_2}A \delta_{\e_1}A )tr_{z^\prime}(F^2) +  2 tr_z (\delta_{\e_2} A
F)tr_{z^\prime}(\delta_{\e_1} AF) ] \quad .
\ee 
Consequently, eq. (\ref{d=6globalanomaly})  is to be modified by terms cubic in the 
gaugini, and the complete result, written in component notation, is finally
\beq {\cal{A}}^{(6)}_\e & & =-\frac{1}{4}\e^{\m\n\a\b\g\delta} c_r^z c^{r z^\prime} tr_z (
\delta_\e A_\m A_\n ) tr_{z^\prime} ( F^\prime_{\a\b}F^\prime_{\g\delta} ) \nonumber\\ &
& -\frac{1}{6}\e^{\m\n\a\b\g\delta} c_r^z c^{rz^\prime}tr_z ( \delta_\e A_\m F_{\n\a}
)\w^{z^\prime}_{\b\g\delta} \nonumber\\ & & +A c_r^z c^{r z^\prime} tr_z (\delta_\e A_\m
F_{\n\r} ) tr_{z^\prime} ( \bar{\l}^\prime \g^{\m\n\r} \l^\prime ) \nonumber\\ & & + B
c_r^z c^{r z^\prime}tr_z ( \delta_\e A_\m \bar{\l})
\g^{\m\n\r} tr_{z^\prime} (\l^\prime F^\prime_{\n\r} ) \nonumber\\ & & + C c_r^z c^{r
z^\prime } tr_z (\delta_\e A_\m \bar{\l}  )
\g_\n  tr_{z^\prime} (\l^\prime F^{\prime \m\n } )\quad ,
\label{completesusyanomaly}
\eeq where the coefficients $A$, $B$ and $C$ satisfy the relations
\beq & & A+B =i \quad, \nonumber\\ & & C=4A - 2B \quad . \label{relations}
\eeq These leave one undetermined parameter, in agreement with the well-known fact that
anomalies are defined up to the variation of local functionals. Indeed, adding to the 
supersymmetry anomaly the term
\be
\delta_\e [(\bar{\l}\g^\a \l^\prime )(\bar{\l} \g_\a \l^\prime )]
\ee corresponds to adding terms like the last three in eq.  (\ref{completesusyanomaly})
with coefficients satisfying  the relations $A+B=0$ and $C=4A-2B$, that thus preserve
eqs. (\ref{relations}).  One can then show that the last of eqs.  (\ref{wesszumino})
generates terms containing one derivative and   four gaugini, that cancel using the
Dirac equation $\g^\m D_\m \l =0 $. Naturally, something similar also happens in
six-dimensional supergravity,  as we are about to verify.

Returning to the supersymmetry anomaly of eq. (\ref{theanomaly}), one can observe  that
the coefficients of the third, fourth  and fifth terms are consistent with eqs. 
(\ref{relations}). Moreover, demanding that the last of  eqs. (\ref{wesszumino}) be
satisfied  fixes the other gauge-invariant terms to give exactly the anomaly in  eq.
(\ref{theanomaly}). Finally, the Wess-Zumino condition is satisfied only  on-shell, and
one obtains
\beq & & (\delta_{\e_1}{\cal{A}}_{\e_2} -\delta_{\e_2}{\cal{A}}_{\e_1}) = {\cal
A}_{\tilde{\e}} + {\cal A}_{\tilde{\L}}  + c_r^z c^{r z^\prime} tr_{z,z^{\prime}}
 \lbrace
- \frac{e}{8}\xi_\sigma ( \bar{\l} \g_\m \l^\prime ) (\bar{\l}\g^\m
\g^\sigma [eq.\l^\prime ]_{(\a = 0) }) \nonumber\\ & & +\frac{e}{16} 
\xi_{i \sigma \delta \tau} \lbrace [\bar{\l}\g^\tau
\l^\prime ]_i (\bar{\l}\g^{\sigma\delta} [eq.\l^\prime ]_{(\a = 0) })+ [\bar{\l}\g^\tau
\l ]_i (\bar{\l}^\prime \g^{\sigma\delta} [ eq. \l^\prime ]_{(\a = 0) }) \rbrace
\rbrace  \quad ,\label{openwz}
\eeq where we have stressed that the corresponding field equation for the gaugini  is
determined by eq. (\ref{completelag}) with $\a = 0$.  To reiterate, the anomaly obtained
for $\a=0$ naturally closes on the corresponding field equation for $\l$. Still, the
identity
\beq && c_r^z c^{r z^\prime} tr_{z,z^{\prime}} \lbrace 
-\frac{e}{8} \xi_\sigma ( \bar{\l} \g_\m \l^\prime ) (\bar{\l}\g^\m
\g^\sigma [eq.\l^\prime ]_{(\a) }) +\frac{e}{16} \xi_{i \sigma \delta \tau}
 \lbrace [\bar{\l}\g^\tau
\l^\prime ]_i (\bar{\l}\g^{\sigma\delta} [eq.\l^\prime ]_{(\a) })\nonumber\\ &&+
[\bar{\l}\g^\tau
\l ]_i (\bar{\l}^\prime \g^{\sigma\delta} [ eq. \l^\prime ]_{(\a) }) \rbrace  \rbrace =
c_r^z c^{r z^\prime} tr_{z,z^{\prime}} \lbrace 
-\frac{e}{8} \xi_\sigma ( \bar{\l} \g_\m \l^\prime ) (\bar{\l}\g^\m
\g^\sigma [eq.\l^\prime ]_{(\a = 0) }) \nonumber\\ & & +\frac{e}{16} 
\xi_{i \sigma \delta \tau} \lbrace [\bar{\l}\g^\tau
\l^\prime ]_i (\bar{\l}\g^{\sigma\delta} [eq.\l^\prime ]_{(\a = 0) })+ [\bar{\l}\g^\tau
\l ]_i (\bar{\l}^\prime \g^{\sigma\delta} [ eq. \l^\prime ]_{(\a = 0) }) \rbrace \rbrace
\nonumber\\ && +
\frac{ie \a}{8} 
\frac{c_r^z c^{r z^\prime}c_s^{z^\prime}c^{s z^{\prime\prime}}}{v_t c^{t z^\prime}}
\xi_i^{\m\n\r}
tr_{z,z^\prime ,z^{\prime\prime} }[\bar{\l}\g_\m \l ]_i 
[\bar{\l}^\prime \g_\n
\l^\prime ]_j [\bar{\l}^{\prime\prime}\g_\r
\l^{\prime\prime} ]_j 
\eeq implies that the last term should somehow be generated in the anomaly,  if the
Wess-Zumino condition is to close for any value of $\a$.  In the presence of ${\cal
L}_{\l^4}$, however, the anomaly is modified by eq. (\ref{extraanomaly}), and applying
the last of eqs. (\ref{wesszumino}) to  this term gives
\beq [\delta_{\e_1},\delta_{\e_2}]{\cal{L}}_{\l^4} & & = ([ \delta_{\e_1},\delta_{\e_2} 
] e) {\a
\over 2} c_r^z c^{r z^\prime } tr_{z, z^\prime} [(\bar{\l} \g_\a 
\l^\prime )(\bar{\l} \g^\a \l^\prime )] \nonumber\\ & & + 2e \a c_r^z c^{r
z^\prime}tr_{z,z^\prime}[(\bar{\l}\g_\a \l^\prime )(\bar{\l}\g^\a 
[\delta_{\e_1},\delta_{\e_2}]\l^\prime )] \quad . \label{extrawz}
\eeq 
The commutator in eq. (\ref{extrawz}) is fully known: in particular, 
the coordinate transformation
in the second term  combines with the commutator on $e$ to give a total  divergence,
while gauge and local Lorentz transformations give a vanishing  result. Moreover, the
field equation is obtained from eq.  (\ref{completelag}). The charge in eq. 
(\ref{centralcharge}) thus plays a crucial role: it generates in eq. (\ref{extrawz}) 
precisely
\be
\frac{i e \a}{8}
\frac{c_r^z c^{r z^\prime}c_s^{z^\prime}c^{s z^{\prime\prime}}}{v_t c^{t z^\prime}}
\xi_i^{\m\n\r} tr_{z,z^\prime ,z^{\prime\prime} } [\bar{\l}\g_\m \l ]_i 
[\bar{\l}^\prime \g_\n
\l^\prime ]_j [\bar{\l}^{\prime\prime}\g_\r
\l^{\prime\prime} ]_j \quad ,
\ee as needed for consistency. Thus, one can understand the rationale behind the
occurrence of the extension  in the  algebra on the gaugini: it lets the Wess-Zumino
conditions close precisely on  the field equations determined by the algebra. Since  the
Wess-Zumino conditions close on the equation of the gaugini,  only these fields perceive
the additional transformation.

\section{Discussion}

In the previous Sections we have completed the coupling  of $(1,0)$  six-dimensional
supergravity to tensor and vector multiplets.  The coupling to tensor multiplets only,
initiated by Romans
\cite{romans},  is of a more conventional nature, and parallels similar constructions in
other supergravity models.  One would expect similar results for the completion of the
$(2,0)$ models in \cite{romans}.  Our work is here  confined to the  field
equations, but a lagrangian formulation of the (anti)self-dual two-forms  is now
possible, following the proposal of Pasti, Sorokin and Tonin 
\cite{pst} and indeed, while this work was being typed, results to this effect have been
presented  in a superspace formulation in \cite{padova}.  On the other hand, the
coupling  to vector multiplets \cite{as}, originally suggested by perturbative type-I
vacua \cite{bs}, is of a more unconventional nature, since it is induced by the residual
anomaly polynomial left over after tadpole conditions are imposed,
\be I_8 = \ - \ \sum_{x,y} \ c^r_x \ c^s_y \ \eta_{rs} \ {\rm tr}_x F^2 \ {\rm tr}_y F^2
\quad .
\ee The corresponding Chern-Simons couplings of the two-forms, 
\be H^r = dB^r - c^{rz} \w_z \quad ,
\ee involve the constants $c^r_z$ and determine related couplings of the other fields. In
particular, the Yang-Mills currents are not conserved, and the  consistent residual
gauge anomaly is accompanied by a corresponding anomaly  in the supersymmetry current
\cite{fms}.  In completing these results to all orders in the fermi fields, we have come
to terms with another peculiar feature of anomalies, neatly displayed by these
``classical'' field equations: anomalous divergences of gauge currents are typically
accompanied by corresponding anomalies in current commutators \cite{anom}. Indeed, we
have discovered an amusing extension of the supersymmetry algebra on  the gaugini, and
we have linked its presence to an ambiguity in the definition of the supergravity model
via Wess-Zumino consistency conditions.   Whereas typical supergravity constructions
yield a unique result, here one is  free to add to the theory a quartic coupling for the
gaugini
\be {\cal{L}}_{\l^4} =  \frac{e \a}{2} \ c_r^z c^{r z^\prime} tr_{z, z^\prime}
[(\bar{\l} \g^\a
\l^\prime )(\bar{\l}\g_\a \l^\prime ) ] \quad ,
\ee whose presence affects only the supersymmetry anomaly.  The Wess-Zumino conditions
for six-dimensional supergravity close only on the field equation of the gaugini, and
are consistent with any choice of $\a$ only thanks to the presence of the extension, as
discussed in Section 3.3. Finally, we should mention that the singular gauge couplings
$v^r c_r^z$ of ref. \cite{as} are accompanied by corresponding divergent fermionic
couplings, partly anticipated by Nishino and Sezgin \cite{ns2}.
 
\vskip 24pt
\section*{Acknowledgment}

We are grateful to Ya.S. Stanev for several stimulating discussions. The work of S.F. was
supported in part by  I.N.F.N., by  DOE grant DE-FG03-91ER40662, Task C. and by E.E.C.
Science Program SC1*CT92-0789.
\vskip 36pt

\section{Appendix: Notations and Conventions}

In six dimensions, a 3-form $X_{\m\n\r}$ is (anti)self-dual if  
\be X_{\m\n\r}=\ (-) \ \frac{1}{6e}\e_{\m\n\r\a\b\g}X^{\a\b\g}\quad .
\ee If $X_{\m\n\r}$ and $Y_{\m\n\r}$ are both (anti)self-dual, 
\be X_{\m\n\r}Y^{\m\n\r}=0
\ee and
\be X_{\m\n\a}Y^{\m\n}{}_\b - X_{\m\n\b}Y^{\m\n}{}_\a =0\quad ,
\ee while if they have opposite duality properties 
\be X_{\m\n\a}Y^{\m\n}{}_\b +X_{\m\n\b}Y^{\m\n}{}_\a =\frac{1}{3}g_{\a\b}
X_{\m\n\r}Y^{\m\n\r}\quad .
\ee Moreover, an (anti)self-dual antisymmetric tensor $X_{\m\n\r}$ satisfies
\be X^{\m\n\r}X_{\a\b\r} = \frac{1}{4} [ -\delta^\m_\b X_{\a\g\delta} X^{\n\g\delta}
+\delta^\n_\b X_{\a\g\delta}X^{\m\g\delta} +
\delta^\m_\a X_{\b\g\delta}X^{\n\g\delta} -\delta^\n_\a X_{\b\g\delta} X^{\m\g\delta } ]
\quad .
\ee

The signature of our metric is $(+,-,...,-)$, and the covariant derivative $D_\m$
contains the full spin connection $\w$, the torsionless Christoffel connection, the
gauge connection and the composite $SO(n)$ connection
\be
S_\m^{mn} = ( \de_\m x^{m}_r ) x^{nr} \quad . 
\ee
The
Clifford  algebra is generated by $\lbrace \g_a ,\g_b \rbrace =2 \eta_{ab}$, and the
chirality matrix is
\be
\g_7 =\g^0 \g^1 \g^2 \g^3 \g^4 \g^5 \quad .
\ee Using $\e^{012345}=+1 $, one obtains
\be
\g^{\m_1 ... \m_n}=- \frac{(-1)^{[n/2]}}{(6-n)!e}\e^{\m_1 ...\m_n \n_1
...\n_{6-n}}\g_{\n_1 ...\n_{6-n}} \g_7 \quad .\label{gammamatrices}
\ee In particular, eq. (\ref{gammamatrices}) shows that $\g_{\m\n\r} \Psi$ is self-dual
if $\Psi$ is  left-handed, and antiself-dual if $\Psi$ is right-handed. 

Spinors are $Sp(2)$ doublets satisfying the symplectic Majorana condition
\be
\Psi^a =\e^{ab} C \bar{\Psi}^T_b \quad,\label{majoranacond}
\ee where
\be
\bar{\Psi}_a =(\Psi^a )^\dagger \g_0
\ee and $\e^{12} =\e_{12}=1$.  The charge conjugation matrix is defined by
\be C\g^\m C = - \g^{\m ,T} \quad,
\ee where $\g^0$ is hermitian and the $\g^i$ are anti-hermitian. Any bilinear
$\bar{\Psi}_a
\chi^b$ carries a pair of $Sp(2)$ indices, and can be decomposed in terms of the
identity and of the three Pauli matrices. Indeed, one can form the bilinears
\be (\bar{\Psi} \chi )=\bar{\Psi}_a \chi^a \quad ,\qquad [\bar{\Psi} \chi ]_i =
\sigma_{i a}{}^b \bar{\Psi}_b \chi^a \quad,
\ee and standard properties imply that
\be
\bar{\Psi}_a \chi^b =\frac{1}{2}\delta_a^b (\bar{\Psi} \chi ) +\frac{1}{2}\sigma_{i
a}{}^b [\bar{\Psi} \chi ]_i\quad.\label{sp2exp}
\ee Using eq. (\ref{majoranacond}), one can then see that the fermi bilinear
$(\bar{\Psi} \chi )$ has standard behavior under Majorana-flip, namely
\be (\bar{\Psi} \chi ) =(\bar{\chi} \Psi ) \quad ,
\ee while all three bilinears $[\bar{\Psi} \chi ]_i $ have the anomalous behavior
\be [\bar{\Psi} \chi ]_i = -[\bar{\chi} \Psi ]_i \quad .
\ee Corresponding relations hold for all fermi bilinears, that naturally display pairs
of opposite behaviors under Majorana flip.  In particular, these properties imply that
\be [ \bar{\Psi} \g_{\m\n\r} \Psi ]_i =0\quad ,
\ee a relation often used in deriving our results.

One can study Fierz relations between spinor bilinears  using eq. (\ref{sp2exp}). If
$\Psi$ and
$\chi$ have the same chirality
\be
\Psi^a \bar{\chi}_b = -\frac{1}{4}\bar{\chi}_b \g^a \Psi^a \g_\a +\frac{1}{48}
\bar{\chi}_b
\g^{\a\b\g} \Psi^a \g_{\a\b\g}\quad ,\label{fierz1}
\ee while if they have opposite chirality
\be
\Psi^a \bar{\chi}_b = -\frac{1}{4} \bar{\chi}_b \Psi^a  +\frac{1}{8} 
\bar{\chi}_b \g^{\a\b} \Psi^a \g_{\a\b} \quad .
\ee 
Interesting results obtain when one (anti)symmetrizes these relations. In
particular, eq. (\ref{fierz1}) implies
\be
\Psi^a \bar{\chi}_b -\chi^a \bar{\Psi}_b = -\frac{1}{4} (\bar{\chi} \g^\a \Psi
)\delta_b^a \g_\a +\frac{1}{48}  [\bar{\chi} \g^{\a\b\g} \Psi ]_i \sigma_{i b}{}^a
\g_{\a\b\g}\quad ,
\ee 
a result often used in the paper.

\end{document}